\documentclass{article} 
\def\version{October 9, 2012}
\usepackage[tbtags]{amsmath}
\usepackage{amsthm}
\usepackage{graphicx}
\usepackage{amssymb,amsfonts,latexsym,times}
\usepackage[latin1]{inputenc}
\usepackage[english]{babel}
\usepackage{bm}
\usepackage{upgreek}
\usepackage{mathrsfs}
\usepackage{color}  

\nonstopmode

\numberwithin{equation}{section}

\DeclareSymbolFont{AMSb}{U}{msb}{m}{n}
\DeclareSymbolFontAlphabet{\mathbb}{AMSb}

\DeclareSymbolFont{EUR}{U}{eur}{m}{n}
\SetSymbolFont{EUR}{bold}{U}{eur}{b}{n}
\DeclareSymbolFontAlphabet{\eur}{EUR}

\DeclareSymbolFont{EUB}{U}{eur}{b}{n}
\SetSymbolFont{EUB}{bold}{U}{eur}{b}{n}
\DeclareSymbolFontAlphabet{\eub}{EUB}

\newcommand{\R}{\mathbb{R}}
\newcommand{\C}{\mathbb{C}}

\newcommand{\abs}[1]{| #1|}
\newcommand{\Abs}[1]{\biggr| #1\biggr|}
\newcommand{\p}{\partial}

\newcommand{\att}[1]{\vert\sb{\sb{#1}}}

\newcommand{\norm}[1]{\Vert #1 \Vert}

\def\spec{\sigma}

\renewcommand{\Re}{\mathop{\rm{R\hskip -1pt e}}\nolimits}
\renewcommand{\Im}{\mathop{\rm{I\hskip -1pt m}}\nolimits}

\DeclareMathSymbol{\sblacksquare}{\mathord}{AMSa}{"04}
\newcommand\blackdiamond{
{\scriptscriptstyle\blacklozenge}
}
\renewcommand\diamond{
{\scriptscriptstyle\lozenge}
}
\theoremstyle{plain}
\newtheorem{lemma}{Lemma}[section]

\theoremstyle{definition}
\newtheorem{definition}[lemma]{Definition}

\theoremstyle{remark}
\newtheorem{remark}[lemma]{Remark}

\begin{document}

\title{On spectral stability of solitary waves of nonlinear Dirac equation
on a line}




\author{
{\sc Gregory Berkolaiko}
\footnote{
Supported in part
by the National Science Foundation under Grant DMS-0907968.
}
\\
{\it\small Texas A\&M University, College Station, TX 77843, U.S.A.}
\\ \\
{\sc Andrew Comech}
\\
{\it\small Texas A\&M University, College Station, TX 77843, U.S.A.}
\\
{\it\small Institute for Information Transmission Problems,
Moscow 101447, Russia}
}

\date{\version}

\maketitle

\begin{abstract}
We study the spectral stability
of solitary wave solutions
to the nonlinear Dirac equation
in one dimension.
We focus on the Dirac equation
with cubic nonlinearity,
known as the Soler model in (1+1) dimensions
and also as the massive Gross-Neveu model.
Presented numerical computations
of the spectrum of linearization
at a solitary wave
show that the solitary waves are spectrally stable.
We corroborate our results
by finding explicit expressions
for several of the eigenfunctions.
Some of the analytic results
hold for the nonlinear Dirac equation
with generic nonlinearity.
\end{abstract}





\section{Introduction}

The study of stability of localized solutions
to nonlinear dispersive equations
takes its origin in
\cite{MR0174304},
where the instability of stationary localized solutions
to nonlinear Klein-Gordon equation
was proved.
It was suggested there that
quasistationary
finite energy
solutions
of the form $\phi(x)e^{-i\omega t}$,
which we call \emph{solitary waves},
could be stable.
The first results
on (spectral) stability of the linearization
at solitary wave solutions
to a nonlinear Schr\"odinger equation
were obtained
in \cite{zakharov-1967,VaKo}.
Orbital stability
and instability
of solitary waves in nonlinear Schr\"odinger and Klein-Gordon
equations
have been extensively studied
in \cite{MR723756,MR804458,MR792821,MR820338,MR901236}.
The asymptotic stability of solitary waves
in nonlinear Schr\"odinger equation
was proved in certain cases in
\cite{MR783974,MR1199635e,MR1170476,MR1681113,MR1835384,MR1972870,MR2443298}.

Systems with Hamiltonians that are not sign-definite
are notoriously difficult,
due to the absence of the a priori bounds
on the Sobolev norm.
Important examples of such systems are
the Dirac-Maxwell system \cite{MR0190520}
and the nonlinear Dirac equation \cite{PhysRevD.1.2766},
which have been
receiving a lot of attention in theoretical physics
in relation to classical models of elementary particles.
The stability
of solitary wave solutions
to the nonlinear Dirac equation
is far from being understood.
Some partial results on the numerical analysis
of spectral stability of solitary waves
are contained in \cite{chugunova-thesis}.
Generalizing
the results on orbital stability of solitary waves
\cite{MR901236}
to the nonlinear Dirac equation
does not seem realistic,
because of the corresponding energy functional being sign-indefinite;
instead, one hopes to prove the asymptotic stability,
using linear stability combined with the dispersive estimates.
The first results on asymptotic stability
for the nonlinear Dirac equation
are already appearing~\cite{2010arXiv1008.4514P,2011arXiv1103.4452B},
with
the assumptions on the spectrum of the linearized equation
playing a crucial role.
In view of these applications,
the spectrum of the linearization at a solitary wave
is of great interest.

In the present paper, we give
numerical and analytical
justifications
of spectral stability
of small amplitude solitary wave
solutions to the nonlinear Dirac
equation in one dimension.

\medskip

Let us remind the terminology.
Given a solitary wave
$\phi(x)e^{-i\omega t}$,
we consider a small perturbation 
of the form
$\psi(x,t)=(\phi(x)+\rho(x,t))e^{-i\omega t}$.
We call a solitary wave
$\phi(x)e^{-i\omega t}$
\emph{linearly unstable}
if the equation on $A$
is given by $\p\sb t\rho=A\rho+o(\rho)$,
with $A$
having eigenvalues with positive real part.
If the entire spectrum of $A$ is on the imaginary axis,
we call the solitary wave
\emph{spectrally stable}.
The solitary wave
is called \emph{orbitally stable}
\cite{MR901236}
if any solution $\psi(t)$
initially close to $\phi$
(in a certain norm,
usually the energy norm)
will exist globally,
remaining close to the orbit spanned by $\phi$ for all times:

\begin{verse}

\qquad\quad
\it
For any $\epsilon>0$
there is $\delta>0$
such that if
$\norm{\psi\sb 0-\phi}<\delta$,
then
there is a solution $\psi(t)$

\qquad\quad
\ \ which exists for all $t\ge 0$
and satisfies
$\psi\att{t=0}=\psi\sb 0$,
\quad
$\sup\limits\sb{t\ge 0}\inf\limits\sb{s\in\R}
\norm{\psi\att{t}-e^{is}\phi}<\epsilon$.
\end{verse}
\noindent
Otherwise, the solitary wave
is called \emph{orbitally unstable}.
A solitary wave is called \emph{asymptotically stable}
if any solution initially close to it
(in a certain norm)
will converge (in a certain norm) to this or to a nearby
solitary wave.
Linear instability of solitary waves
generically leads to \emph{orbital instability}
\cite{MR948770,2010arXiv1009.5184G};
at the same time,
\emph{spectral stability}
does not imply neither orbital nor asymptotic stability.

\medskip

\noindent
{ACKNOWLEDGMENTS}
The authors would like to thank
Marina Chugunova
for providing us with her preliminary numerical results
on spectral properties
of coupled mode equations
(see also \cite{chugunova-thesis})
which greatly stimulated our research.
The authors are grateful
to
Nabile Boussaid,
Thomas Chen,
Linh Nguyen,
Dmitry Pelinovsky,
Bjorn Sandstede,
Walter Strauss,
Boris Vainberg,
and Michael Weinstein
for most helpful discussions.

\section{Nonlinear Dirac equation}

The nonlinear Dirac equation
has the form
\begin{equation}\label{dirac}
i\p\sb t\psi=-i\sum\sb{j=1}\sp{n}\alpha\sb{j}\p\sb{x\sb{j}}\psi
+\beta g(\psi\sp\ast\beta\psi)\psi,
\qquad
\psi(x,t)\in\mathbb{C}^N,
\qquad x\in\R^n,
\end{equation}
with
$\psi\sp\ast$
being the Hermitian conjugate of $\psi$.
The Hermitian matrices $\alpha\sb{j}$ and $\beta$
are chosen so that
\[
\alpha\sb{j}^2=I,
\qquad
\beta^2=I,
\qquad
\{\alpha\sb{j},\alpha\sb k\}=2\delta\sb{j k},
\qquad
\{\alpha\sb{j},\beta\}=0,
\qquad
1\le j,\,k\le n.
\]
We assume that the nonlinearity $g$
is smooth and real-valued.
We denote $m\equiv g(0)$.
Equation \eqref{dirac}
with $n=3$ and $g(s)=1-s$
is known as the Soler model \cite{PhysRevD.1.2766}
(when $n=3$, one can take
Dirac spinors  with $N=4$ components).
The case $n=1$
(when one can take spinors with $N=2$ components)
is known as the massive Gross-Neveu model
\cite{PhysRevD.10.3235,PhysRevD.12.3880}.

In the present paper, we consider the Dirac equation in $\R^1$:
\begin{equation}\label{nld-1d}
i(\p\sb t+\alpha\p\sb x)\psi =g(\psi\sp\ast\beta\psi)\beta\psi,
\qquad
\psi(x,t)\in\C^2,\qquad
x\in\R^1.
\end{equation}
As $\alpha$ and $\beta$,
we choose
\begin{equation}\label{ab}
\alpha =-\sigma\sb 2,
\qquad
\beta=\sigma\sb 3,
\end{equation}
with the Pauli matrices
$
\sigma\sb 1=\left(\begin{matrix}0&1\\1&0\end{matrix}\right),
$
$
\sigma\sb 2=\left(\begin{matrix}0&-i\\i&0\end{matrix}\right),
$
$
\sigma\sb 3
=\left(\begin{matrix}1&0\\0&-1\end{matrix}\right).
$
Noting that
$\psi\sp\ast\sigma\sb3\psi=\abs{\psi\sb1}^2-\abs{\psi\sb 2}^2$, 
we rewrite equation~\eqref{nld-1d} as the following system:
\begin{equation}\label{nld-2c}
\left\{
\begin{array}{ll}
i\p\sb t \psi\sb 1=\p\sb x \psi\sb 2+g(\abs{\psi\sb
1}^2-\abs{\psi\sb 2}^2)\psi\sb 1,
\\
i\p\sb t \psi\sb 2=-\p\sb x \psi\sb 1-g(\abs{\psi\sb
1}^2-\abs{\psi\sb 2}^2)\psi\sb 2.
\end{array}
\right.
\end{equation}

\section{Solitary wave solutions}


We start by demonstrating the existence of solitary wave solutions and
exploring their properties.

\begin{definition}
The solitary waves are solutions
to \eqref{dirac}
of the form
\[
\psi(x,t)=\phi\sb\omega(x)e^{-i\omega t}, \qquad
\phi\sb\omega\in H\sp 1(\R^n,\C^N),\ \omega\in\R.
\]
\end{definition}

The following result follows from \cite{MR847126}.

\begin{lemma}\label{lemma-existence-nld-1d}
Assume that
\begin{equation}\label{f-positive}
m:=g(0)>0.
\end{equation}
Let $G(s)$ be the antiderivative of $g(s)$ such that $G(0)=0$.
Assume that
for given $\omega\in\R$, $0<\omega<m$, there exists
$\mathscr{X}\sb\omega>0$ such that
\begin{equation}\label{def-Xi}
\omega\mathscr{X}\sb\omega=G(\mathscr{X}\sb\omega), \qquad
\omega\ne g(\mathscr{X}\sb\omega),
\qquad\mbox{and}\qquad
\omega s<G(s)\quad{\rm for}
\ s\in(0,\mathscr{X}\sb\omega).
\end{equation}
Then there is a solitary wave solution
$\psi(x,t)=\phi\sb\omega(x)e^{-i\omega t}$,
where
\begin{equation}\label{psi-v-u}
\phi\sb\omega(x)
=\left[\begin{matrix}v(x)\\u(x)\end{matrix}\right],
\qquad v,\ u\in H\sp 1(\R),
\end{equation}
with both $v$ and $u$ real-valued,
$v$ being even and $u$ odd.

More precisely, let us define $\mathscr{X}(x)$ and $\mathscr{Y}(x)$ by
\begin{equation}\label{def-xi-eta}
\mathscr{X}=v^2-u^2,\qquad \mathscr{Y}=vu.
\end{equation}
Then $\mathscr{X}(x)$ is the solution to
\begin{equation}
\mathscr{X}''=-\p\sb\mathscr{X} (-2G(\mathscr{X})^2+2\omega^2\mathscr{X}^2), \qquad
\mathscr{X}(0)=\mathscr{X}\sb\omega,\qquad \mathscr{X}'(0)=0,
\end{equation}
and $\mathscr{Y}(x)=-\frac{1}{4\omega}\mathscr{X}'(x)$.
\end{lemma}

\begin{proof}
  From (\ref{nld-2c}), we obtain:
\begin{equation}\label{omega-v-u}
\left\{
\begin{array}{ll}
\omega v=\p\sb x u+g(\abs{v}^2-\abs{u}^2)v,
\\
\omega u=-\p\sb x v-g(\abs{v}^2-\abs{u}^2)u.
\end{array}
\right.
\end{equation}
Assuming that both $v$ and $u$ are real-valued
(this will be justified once we found
real-valued $v$ and $u$),
we can rewrite (\ref{omega-v-u}) as the following
Hamiltonian system, with $x$ playing the role of time:
\begin{equation}\label{stat-eqn}
\left\{
\begin{array}{ll}
\p\sb x u=\omega v-g(v^2-u^2)v=\p\sb v h(v,u),
\\
-\p\sb x v=\omega u+g(v^2-u^2)u=\p\sb u h(v,u),
\end{array}
\right.
\end{equation}
where the Hamiltonian $h(v,u)$ is given by
\begin{equation}\label{def-h}
h(v,u)=\frac{\omega}{2}(v^2+u^2)-\frac{1}{2}G(v^2-u^2).
\end{equation}
The solitary wave corresponds to a trajectory of this Hamiltonian
system such that
\[
\lim\limits\sb{x\to\pm\infty}v(x)
=\lim\limits\sb{x\to\pm\infty}u(x)=0,
\]
hence
$\lim\limits\sb{x\to\pm\infty}\mathscr{X}=0$.
Since $G(s)$ satisfies
$G(0)=0$,
we conclude that
\begin{equation}\label{h-is-0}
h(v(x),u(x))\equiv 0,
\end{equation}
which leads to
\begin{equation}\label{phi-phi-g}
\omega(v^2+u^2)=G(v^2-u^2).
\end{equation}
Studying the level curves which solve this equation
is most convenient in the coordinates
\[
\mathscr{X}=v^2-u^2,
\qquad
\mathscr{Z}=v^2+u^2;
\]
see Figure~\ref{fig-solitons-existence}.
We conclude from
\eqref{phi-phi-g}
and
Figure~\ref{fig-solitons-existence}
that
solitary waves may correspond to $\abs{\omega}<m$,
$\omega\ne 0$.

\begin{remark}
If $\omega>0$,
then there are solitary waves
such that $v$ is nonzero
while $u$ changes its sign
(shifting the origin,
we may assume that this happens at $x=0$).
For $\omega<0$,
there are solitary waves
such that $u\ne 0$,
while $v$ changes its sign.
\end{remark}

\begin{remark}
In the case when $G(s)$ is odd,
for each solitary wave corresponding to $\omega\in\R$
there is a solitary wave corresponding to $-\omega$.
More precisely,
in this case,
if
$\left[\begin{matrix}v(x)\\u(x)\end{matrix}\right]e^{-i\omega t}$
is a solitary wave,
then so is
$\left[\begin{matrix}u(x)\\v(x)\end{matrix}\right]e^{i\omega t}$.
\end{remark}

\begin{figure}[htbp]
\hskip 4cm
\includegraphics{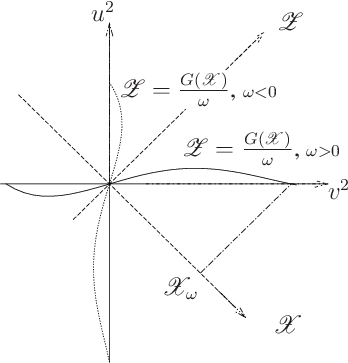}
\caption{\footnotesize Existence of solitary waves
in the coordinates
$\mathscr{X}=v^2-u^2$, $\mathscr{Z}=v^2+u^2$.
Solitons
with $\omega>0$
and $\omega<0$
correspond to the bump
on the $v^2$ axis
and
to the dotted bump
on the $u^2$ axis
(respectively)
in the first quadrant.}
\label{fig-solitons-existence}
\end{figure}

The functions $\mathscr{X}(x)$ and $\mathscr{Y}(x)$ introduced in
(\ref{def-xi-eta}) are to solve
\begin{equation}\label{xi-eta-system}
\left\{
\begin{array}{ll}
\mathscr{X}' =-4\omega\mathscr{Y},
\\
\mathscr{Y}' =-(v^2+u^2)g(\mathscr{X})+\omega\mathscr{X}
=-\frac{1}{\omega}G(\mathscr{X})g(\mathscr{X})+\omega\mathscr{X},
\end{array}
\right.
\end{equation}
and to have the asymptotic behavior
$\lim\sb{\abs{x}\to\infty}\mathscr{X}(x)=0$,
$\lim\sb{\abs{x}\to\infty}\mathscr{Y}(x)=0$.
In the second equation in
(\ref{xi-eta-system}), we used the relation (\ref{phi-phi-g}).
The system (\ref{xi-eta-system}) can be
written as the following equation on $\mathscr{X}$:
\begin{equation}\label{xi-p-p}
\mathscr{X}'' = -\p\sb\mathscr{X}(-2 G(\mathscr{X})^2+2\omega^2\mathscr{X}^2).
\end{equation}
This equation describes a particle in the potential
$V\sb\omega(s)=-2 G(s)^2+2\omega^2 s^2$;
see Figure~\ref{f-vs-x}.
Due to the energy
conservation (with $x$ playing the role of time), we get:
\begin{equation}\label{h-is-000}
\frac{\mathscr{X}'^2}{2}-2 G(\mathscr{X})^2+2\omega^2\mathscr{X}^2
=\frac{\mathscr{X}'^2}{2}+V\sb\omega(\mathscr{X})
=0.
\end{equation}
Using
the expression for $\mathscr{X}'$ from
\eqref{xi-eta-system},
relation
\eqref{h-is-000}
could be rewritten as
\begin{equation}\label{h-is-00}
0=
\frac{\mathscr{X}'^2}{2}+V\sb\omega(\mathscr{X}) =8\omega^2\mathscr{Y}^2
-2G(\mathscr{X})^2+2\omega^2\mathscr{X}^2
=2\omega^2\big(4v^2u^2+(v^2-u^2)^2\big)-2G^2,
\end{equation}
which follows from (\ref{phi-phi-g}).

For a particular value of $\omega$, there will be a positive
solution $\mathscr{X}(x)$ such that
$\lim\sb{x\to\pm\infty}\mathscr{X}(x)=0$
if there exists $\mathscr{X}\sb\omega>0$
so that (\ref{def-Xi}) is satisfied (see Figure~\ref{f-vs-x}).
We shift $x$ so that
$\mathscr{X}$ so that $\mathscr{X}(0)=\mathscr{X}\sb\omega$;
then $\mathscr{X}(x)$ is an even function.

\begin{figure}[htbp]
\bigskip
\bigskip
\hskip 4cm
\includegraphics{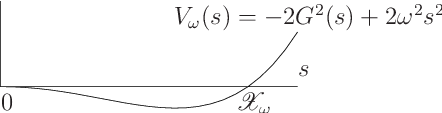}
\caption{\footnotesize Effective potential $V\sb\omega(s)$.
A solitary wave corresponds to a trajectory
which satisfies $\mathscr{X}(0)=\mathscr{X}\sb\omega$,
$\lim\sb{\abs{x}\to\infty}\mathscr{X}(x)=0$.
}
\label{f-vs-x}
\end{figure}

Once $\mathscr{X}(x)$ is known,  $\mathscr{Y}(x)$
is obtained from (\ref{xi-eta-system}),
and then
we can express $v(x)$, $u(x)$.
\end{proof}

\begin{remark}\label{remark-ed}
Note that
for $0<\abs{\omega}<1$,
the functions $v(x)$ and $u(x)$
are exponentially decaying
as $\abs{x}\to\infty$.
Indeed, the exponential decay of
$\mathscr{X}(x)$ could be deduced from
\eqref{xi-p-p}.
Then the exponential decay
of $\mathscr{Z}(x)=v(x)^2+u(x)^2$ follows from
the relation $\mathscr{Z}=G(\mathscr{X})/\omega$
(Cf. \eqref{phi-phi-g}).
\end{remark}

\subsection{Explicit solitary waves in a particular case}
\label{sec:explicit}

As shown in
\cite{PhysRevD.12.3880}
for the massive Gross-Neveu model
(the Soler model in 1D),
in the special case of the potential
\begin{equation}\label{gx-x-x2}
G(s)=s-\frac{s^2}{2},
\end{equation}
the solitary waves can be found explicitly.
Substituting
$G(s)$ from \eqref{gx-x-x2}
into \eqref{h-is-000},
we get the following relation:
\begin{equation}
dx=-\frac{d\mathscr{X}}{2\sqrt{(\mathscr{X}-\mathscr{X}^2/2)^2-\omega^2\mathscr{X}^2}}
=-\frac{d\mathscr{X}}{2\mathscr{X}\sqrt{(1-\mathscr{X}/2)^2-\omega^2}}.
\end{equation}
We use the substitution
\begin{equation}
1-\frac{\mathscr{X}}{2}=\frac{\omega}{\cos{2\Theta}}, \qquad
\mathscr{X}=2\Big(1-\frac{\omega}{\cos{2\Theta}}\Big).
\end{equation}
Then
\begin{equation}
dx =-\frac{d\mathscr{X}}{2\mathscr{X}\sqrt{(1-\mathscr{X}/2)^2-\omega^2}}
=\frac{2\frac{2\omega\sin 2\Theta}{\cos^2{2\Theta}}\,d\Theta}
{4(1-\frac{\omega}{\cos{2\Theta}}) \sqrt{\frac{\omega^2}{\cos^2
2\Theta}-\omega^2}} =\frac{d\Theta}{\cos{2\Theta}-\omega},
\end{equation}
\begin{equation}
x=\frac{1}{2\kappa}\ln\Abs{
\frac{\sqrt{\mu}+\tan\Theta}{\sqrt{\mu}-\tan\Theta}},
\end{equation}
where
\begin{equation}
\label{eq:kappa_mu}
\kappa=\sqrt{1-\omega^2},
\qquad
\mu=\frac{1-\omega}{1+\omega}.
\end{equation}
Then
\begin{equation}\label{Theta-is-small}
(\sqrt{\mu}+\tan\Theta)e^{2\kappa x} =\sqrt{\mu}-\tan\Theta,
\qquad
\tan\Theta(x)=-\sqrt{\mu}\tanh{\kappa x}.
\end{equation}
Also note that
\begin{equation}
\mathscr{X}(x)
=2\left(1-\frac{\omega}{\cos 2\Theta}\right)
=2\left(1-\frac{\omega}{2\cos^2\Theta-1}\right)
=2\left(1-\omega\frac{1+\tan^2\Theta(x)}{1-\tan^2\Theta(x)}\right),
\end{equation}
and then
\begin{eqnarray}
\mathscr{Y}(x)=-\frac{1}{4\omega}\mathscr{X}'(x)
=-\frac{1}{4}\frac{2}{\cos^2 2\Theta}(-2\sin 2\Theta)\frac{d\Theta}{dx}
\nonumber
\\
=-\frac{1}{4}\frac{2}{\cos^2 2\Theta}(-2\sin 2\Theta)
(\cos 2\Theta-\omega)
=\frac{\mathscr{X}}{2}\tan 2\Theta
\nonumber
\\
=
\frac{\mathscr{X}}{2}\frac{2\tan\Theta}{1-\tan^2\Theta}
=
-\mathscr{X}(x)
\frac{\sqrt{\mu}\tanh\kappa x}{1-\mu\tanh^2\kappa x}.
\nonumber
\end{eqnarray}
Denote
\begin{equation}
\mathscr{Z}(x)=v^2(x)+u^2(x).
\end{equation}
Then
\[
\mathscr{Z}(x)
=\frac{2}{\cos 2\Theta(x)}
\left(1-\frac{\omega}{\cos 2\Theta(x)}\right)
=2\frac{1+\tan^2\Theta(x)}{1-\tan^2\Theta(x)}
\left(1-\omega\frac{1+\tan^2\Theta(x)}{1-\tan^2\Theta(x)}\right).
\]
The other functions are expressed from $\mathscr{Z}$ as follows:
\begin{equation}
\label{eq:uv_thru_Z}
v(x)=\sqrt{\mathscr{Z}(x)}\cos\Theta(x),
\qquad u(x)=-\sqrt{\mathscr{Z}(x)}\sin\Theta(x),
\end{equation}
\begin{equation}
\mathscr{X}(x)=\mathscr{Z}(x)\cos 2\Theta(x),
\qquad \mathscr{Y}(x)=-\frac 1 2\mathscr{Z}(x)\sin 2\Theta(x).
\end{equation}
Combining equations (\ref{eq:uv_thru_Z}) with (\ref{eq:kappa_mu}),
(\ref{Theta-is-small})
and using basic trigonometric identities, we obtain the
following explicit formulae for $v(x)$ and $u(x)$:
\begin{equation}
\label{eq:v_and_u}
v(x)= \frac{\sqrt{2(1-\omega)}}{(1-\mu \tanh^2\kappa x)\cosh\kappa x},
\qquad
u(x)= \frac{\sqrt{2\mu(1-\omega)}\tanh\kappa x}{ 
  (1-\mu \tanh^2\kappa x)\cosh\kappa x}.
\end{equation}

\begin{remark}\label{remark-omega-1}
By \eqref{Theta-is-small},
$\tan\Theta$ changes from $\sqrt{\mu}$ to $-\sqrt{\mu}$
as $x$ changes from $-\infty$ to $+\infty$.
Thus, in the limit $\omega\to 1$,
when $\mu\to 0$,
one has
$\mathscr{X}\approx\mathscr{Z}$,
while
$\abs{\mathscr{Y}}\lesssim\mathscr{Z}\sqrt{\mu}$.
\end{remark}


\section{Linearization at a solitary wave}

To analyze the stability of solitary waves we consider the solution in the
form of the Ansatz
\begin{equation}
\psi(x,t)=(\phi\sb\omega(x)+\rho(x,t))e^{-i\omega t},
\qquad
\phi\sb\omega(x)
=\left[\begin{matrix}v(x)\\u(x)\end{matrix}\right]\in\R^2,
\qquad
\rho(x,t)\in\C^2.
\end{equation}
Then, by (\ref{dirac}),
$
i\p\sb t\rho+\omega\rho =-i\sum\sb{j}\alpha\p\sb{x\sb{j}}\rho
+\beta \left[
g((\bar\phi\sb\omega+\bar{\rho})(\phi\sb\omega+\rho))(\phi\sb\omega+\rho)
-g(\bar\phi\sb\omega\phi\sb\omega)\phi\sb\omega\right].
$
The linearized equation on $\rho$ is given by
\[
i\p\sb t\rho=-i\sum\sb{j}\alpha\p\sb{x\sb{j}}\rho -\omega\rho
+\beta \left[ g(\bar\phi\sb\omega\phi\sb\omega)\rho
+(\bar\phi\sb\omega\rho +\bar{\rho}\phi\sb\omega)
g'(\bar\phi\sb\omega\phi\sb\omega)\phi\sb\omega \right].
\]
The linearized equation on
$
\eub{R}(x,t)=\left[
\begin{matrix}
\Re\rho(x,t)
\\
\Im\rho(x,t)
\end{matrix}\right]\in\R^4
$
has the following form:
\begin{equation}\label{l-d}
\p\sb t
\eub{R}
=
\left[
\begin{matrix}0&{I}\sb 2\\-{I}\sb 2&0\end{matrix}\right]
\left[
\begin{matrix}\eur{L}\sb{1}&0\\0&\eur{L}\sb{0}\end{matrix}\right]
\eub{R}
=
\eub{J}\eub{L}\,
\eub{R},
\qquad
\eub{R}(x,t)\in\R^4,
\end{equation}
where
$
\eub{J}=
\left[
\begin{matrix}
0&{I}\sb 2\\-{I}\sb 2&0
\end{matrix}\right],
$
with
${I}\sb 2$ the $2\times 2$ unit matrix,
and
$\eub{L}
=\left[
\begin{matrix}\eur{L}\sb{1}&0\\0&\eur{L}\sb{0}\end{matrix}
\right]$,
where
$\eur{L}\sb{0}$, $\eur{L}\sb{1}$
are self-adjoint operators given by
\[
\eur{L}\sb{0}
=
\left[
\begin{matrix}
g(v^2-u^2)-\omega&\p\sb x
\\
-\p\sb x&-g(v^2-u^2)-\omega
\end{matrix}\right],
\qquad
\eur{L}\sb{1}
=
\eur{L}\sb{0}
+
2g'(v^2-u^2)
\left[
\begin{matrix}
v^2&-vu
\\
-vu&u^2
\end{matrix}\right].
\]

\begin{remark}
The operators $\eub{L}$,
$\eur{L}\sb{0}$, and $\eur{L}\sb{1}$
depend on the parameter $\omega$.
\end{remark}


\section{Spectra of $\eur{L}\sb{0}$ and $\eur{L}\sb{1}$}
\label{sec:spec_Lpm}

While we are ultimately interested in the spectrum of the operator $\eub{J}\eub{L}$,
we start by analyzing the spectra of
$\eur{L}\sb{0}$ and $\eur{L}\sb{1}$,
which are easier to
compute and which will shed some light on the behaviour of the full operator
$\eub{J}\eub{L}$.

\begin{lemma}
The essential spectrum of the operators $\eur{L}\sb{1}$ and $\eur{L}\sb{0}$
is given by
$\R\backslash(-1-\omega,1-\omega)$.
There are no eigenvalues
embedded into the essential spectrum.
\end{lemma}

\begin{proof}
Due to the exponential decay of the solitary wave components $u$ and
$v$
(see Remark~\ref{remark-ed}),
the operators $\eur{L}\sb{0}$, $\eur{L}\sb{1}$ are relatively compact
perturbations of the operator
\begin{equation}
\label{eq:L-infinity-def}
\eur{D}-\omega,
\qquad
\mbox{where}
\quad
\eur{D}
=-i\alpha\p\sb x+\beta
=\left[\begin{matrix}1&\p\sb x\\-\p\sb x&-1\end{matrix}\right].
\end{equation}
The free Dirac operator $\eur{D}$ is a (matrix) differential operator
with constant coefficients.  Its essential spectrum is given by the
values $\lambda$ for which $\ker(\eur{D}-\lambda)$ contains
bounded functions.
In other words, we are looking for solutions of
$(\eur{D}-\lambda)\psi=0$ of the form
$\psi=Z e^{i\xi x}$,
with real $\xi$ and constant $Z\in\R^2$.  Calculating the
determinant of the symbol of $\eur{D}-\lambda$, we get
\begin{equation*}
\det\left[
\begin{matrix}
  1-\lambda& i \xi
  \\
  -i\xi x&-1-\lambda
\end{matrix}
\right]
=\lambda^2 - 1 - \xi^2 = 0.
\end{equation*}
Thus the essential spectrum is the range of
$\lambda=\pm \sqrt{1+\xi^2}$ when $\xi\in\mathbb{R}$,
that is, two intervals
$(-\infty,-1]$ and $[1,\infty)$.

Regarding the embedded eigenvalues, the space of solutions of
$(\eur{L}\sb{0}-\lambda)\Psi=0$
(similarly,
$(\eur{L}\sb{1}-\lambda)\Psi=0$)
is spanned by the two Jost solutions,
i.e. the solutions that have the same asymptotics as the solutions
of $(\eur{D}-\omega -\lambda)\Psi=0$.
For $\lambda\in\sigma_{ess}$, there
are two oscillating Jost solutions which cannot combine to produce a
decaying solution.  Thus there are no eigenfunctions corresponding
to $\lambda\in\sigma_{ess}$.
\end{proof}

The spectrum of $\eur{L}\sb{0}$
has the following symmetry property.

\begin{figure}[htbp]
\includegraphics[width=16cm,height=9cm]{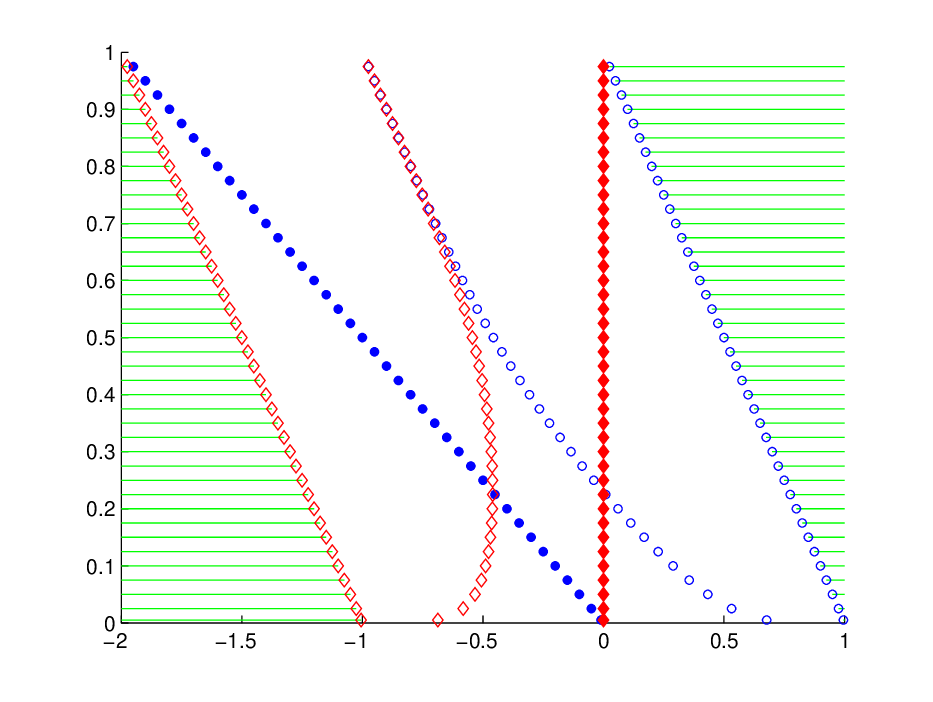}
\caption{\footnotesize $\sigma(\eur{L}\sb{0})$.
Eigenvalues ($\blackdiamond$, $\bullet$;
``even'' and ``odd'', respectively)
and the values of $\lambda$
corresponding to antibound states
($\diamond$, $\circ$;
also ``even'' and ``odd'', respectively).
We say that the eigenvalue is ``even''
(``odd'')
if the first component of the corresponding eigenfunction
is even (odd, respectively).
}
\label{Hminus-spectrum}
\bigskip
\includegraphics[width=16cm,height=9cm]{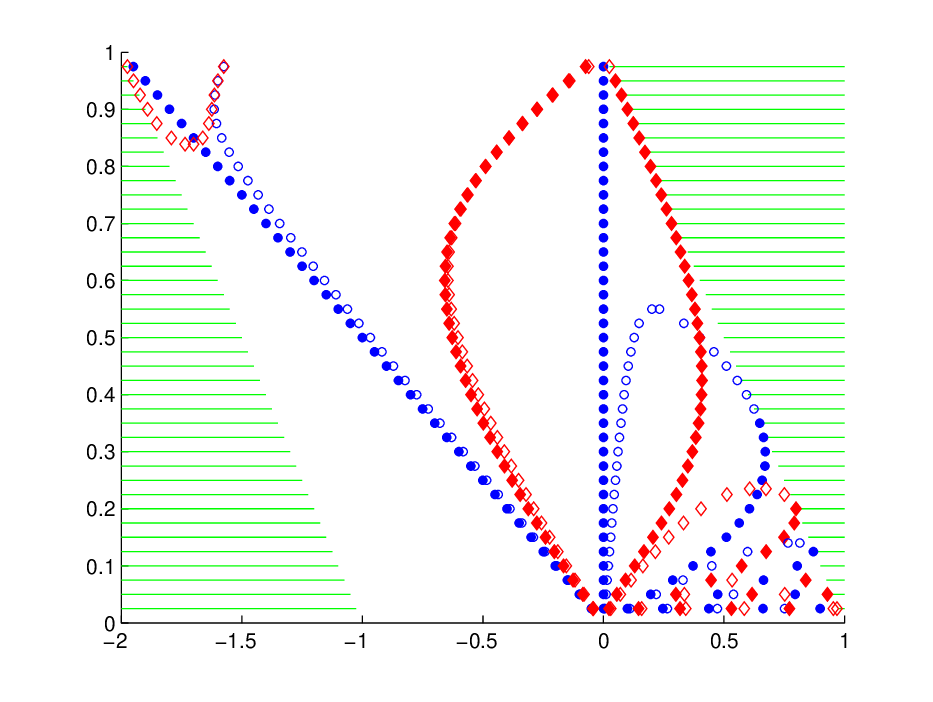}
\caption{\footnotesize 
$\sigma(\eur{L}\sb{1})$.
Eigenvalues (``even'' $\blackdiamond$ and ``odd'' $\bullet$)
and the values of $\lambda$
corresponding to antibound states
(``even'' $\diamond$ and ``odd'' $\circ$).
}
\label{Hplus-spectrum}
\end{figure}

\begin{lemma}\label{lem:sym_of_Lminus}
For each $\omega$
such that there is a solitary wave solution
$\phi(x)e^{-i\omega t}$
to \eqref{nld-1d},
the spectrum of
$\eur{L}\sb{0}$
is symmetric with respect to $\lambda=-\omega$.
See Figure~\ref{Hminus-spectrum}.
\end{lemma}

\begin{proof}
It suffices to check that
if
$\Psi(x)=\left[\begin{matrix}R(x)\\S(x)\end{matrix}\right]\in\C^2$
satisfies $(\eur{L}\sb{0}-\lambda)\Psi=0$,
then
$\Theta(x)=\left[\begin{matrix}S(x)\\R(x)\end{matrix}\right]$
satisfies $(\eur{L}\sb{0}+(2\omega+\lambda))\Theta=0$.
\end{proof}

\begin{remark}
The statement of Lemma~\ref{lem:sym_of_Lminus}
takes place for any nonlinearity $g(s)$.
\end{remark}

Next we explicitly find two eigenvalues together with their
eigenvectors for each of the operators $\eur{L}\sb{0}$, $\eur{L}\sb{1}$.

\begin{lemma}\label{lemma-hm-2omega}
\begin{enumerate}
\item
$\spec\sb d(\eur{L}\sb{0})\supset\{0\}$,
$\spec\sb d(\eur{L}\sb{1})\supset\{0\}$.
The corresponding eigenspaces are given by
\[
\ker\eur{L}\sb{0}=\mathop{\rm Span}\langle\phi\rangle
= \mathop{\rm Span}\left\langle
\left[\begin{matrix}v\\u\end{matrix}\right]
\right\rangle,
\qquad
\ker\eur{L}\sb{1}=\mathop{\rm Span}\langle\p\sb x\phi\rangle
= \mathop{\rm Span}\left\langle
\left[\begin{matrix} v' \\ u' \end{matrix}\right]
\right\rangle.
\]
\item
$\spec\sb{p}(\eur{L}\sb{0})\supset\{-2\omega\}$,
$\spec\sb{p}(\eur{L}\sb{1})\supset\{-2\omega\}$.
The eigenfunction
of both $\eur{L}\sb{0}$ and $\eur{L}\sb{1}$
corresponding to the eigenvalue $-2\omega$
is given by
$\Psi(x)=\left[\begin{matrix}u(x)\\v(x)\end{matrix}\right]$.
\end{enumerate}
\end{lemma}

\begin{proof}
By \eqref{omega-v-u},
$\eur{L}\sb{0}
\left[\begin{matrix}v\\u\end{matrix}\right]=0$, hence
$\phi=\left[\begin{matrix}v\\u\end{matrix}\right]\in\ker\eur{L}\sb{0}$.
Since there are two Jost solutions of $\eur{L}\sb{0}$ corresponding
to $\lambda=0$
with prescribed asymptotic behavior as $x\to +\infty$,
with one growing at $+\infty$ and the other decaying,
there are no more $L\sp 2$ eigenfunctions corresponding to
$\lambda=0$.
For more on Jost solutions,
see Section~\ref{sect-jost}.

  From Lemma~\ref{lem:sym_of_Lminus} we immediately get
that $\lambda=-2\omega$ is an eigenvalue with the
corresponding eigenfunction
given by
$\left[\begin{matrix}u\\v\end{matrix}\right]$.

Turning our attention to the operator $\eur{L}\sb{1}$, we take the
derivative of the relation $\eur{L}\sb{0}\phi=0$ with respect to $x$
to obtain
\[
\eur{L}\sb{1}
\left[\begin{matrix}v'\\u'\end{matrix}\right]
=
\left[
\begin{matrix}
2g'v^2+g-\omega&\p\sb x-2g'vu
\\
-\p\sb x-2g'vu&2g'u^2-g-\omega
\end{matrix}\right]
\left[\begin{matrix}v'\\u'\end{matrix}\right]
=
0.
\]
Using \eqref{omega-v-u} again, we get
\[
(\eur{L}\sb{1}+2\omega)
\left[\begin{matrix}u\\v\end{matrix}\right]
=
\left[
\begin{matrix}
2g'v^2+g+\omega&\p\sb x-2g'vu
\\
-\p\sb x-2g'vu&2g'u^2-g+\omega
\end{matrix}\right]
\left[\begin{matrix}u\\v\end{matrix}\right]
=
\left[
\begin{matrix}
g+\omega&\p\sb x
\\
-\p\sb x&-g+\omega
\end{matrix}\right]
\left[\begin{matrix}u\\v\end{matrix}\right]
=0.
\]
The last equality is due to \eqref{omega-v-u}.
Again, there are no more eigenfunctions
since there is
one Jost solution growing as $x\to+\infty$
and the other decaying,
and one can not use them to construct more than one
eigenfunction.
\end{proof}

At the thresholds
(the endpoints of the essential spectrum)
the solutions of
$(\eur{L}\sb{0}-\lambda)\Psi=0$
and
$(\eur{L}\sb{1}-\lambda)\Psi=0$
are, in general,
linearly growing.
However, the operator $\eur{L}\sb{0}$ has
``resonances'', that is,
generalized eigenfunctions that are uniformly bounded.

\begin{lemma}
\label{lem:explicit_resonances}
For the nonlinearity $g(s)=1-s$
(the Soler model), the values
$\lambda=1-\omega$
and
$\lambda=-1-\omega$
are resonances of $\eur{L}\sb{0}$.
\end{lemma}

\begin{proof}
The generalized eigenfunction
corresponding to $\lambda=1-\omega$
is explicitly given by
\[
\Psi(x)=\left[\begin{matrix}R(x)\\S(x)\end{matrix}\right],
\qquad
\mbox{with}
\quad
R(x)=\frac{u(x)v(x)}{v(x)^2-u(x)^2},
\qquad
S(x)=\frac{v(x)^2-\frac{1+\omega}{1-\omega}u(x)^2}{v(x)^2-u(x)^2}.
\]
By Lemma~\ref{lem:sym_of_Lminus},
the generalized eigenfunction
corresponding to $\lambda=-1-\omega$
is
$\Psi=\left[\begin{matrix}S\\R\end{matrix}\right]$.
\end{proof}

The numerical computations show that
for the nonlinearity $g(s)=1-s$
(the Soler model),
there are no other eigenvalues in
$\eur{L}\sb{0}$; see Figure~\ref{Hminus-spectrum}, bold symbols.
This agrees with \cite{chugunova-thesis}. 
The transparent symbols on
Figure~\ref{Hminus-spectrum}
denote
\emph{antibound} states which will be discussed in detail in
Section~\ref{sec:antibound}.
Note that the antibound states
numerically found at the edges of the essential spectrum are nothing
else but the resonances described in Lemma~\ref{lem:explicit_resonances}.

The spectrum of $\eur{L}\sb{1}$,
besides eigenvalues $\lambda=0$ and $\lambda=-2\omega$
discussed in Lemma~\ref{lemma-hm-2omega},
may contain more eigenvalues.
For the nonlinearity $g(s)=1-s$,
the numerical computation of the spectrum
of $\eur{L}\sb{1}$
is on Figure~\ref{Hplus-spectrum}.
On that picture, eigenvalues are represented by the bold symbols.
There are 4 eigenvalues that belong to the spectrum for
all values of $\omega$ starting from $\omega=1$.
Moreover, there are eigenvalues that
``emerge'' from the essential spectrum as $\omega$ decreases.  In
fact, they can be traced to being antibound states
prior to becoming eigenstates.
We will discuss this in more detail in
Section~\ref{sec:antibound}.

\newpage

\section{Spectrum of $\eub{J}\eub{L}$}

The analysis of the spectrum of $\eub{J}\eub{L}$ builds upon what has
been discovered for the operators $\eur{L}\sb{0}$, $\eur{L}\sb{1}$
in Section~\ref{sec:spec_Lpm}.
In particular, starting from the explicit
eigenvectors for
$\eur{L}\sb{0}$, $\eur{L}\sb{1}$
we are able to construct explicit
eigenvectors for
$\eub{J}\eub{L}$.

\begin{lemma}\label{lemma-spectrum-l}
For any nonlinearity $g(s)$
in \eqref{nld-1d}
with $g(0)=1$,
for linearization
at a solitary wave $\phi(x)e^{-i\omega t}$
with $\abs{\omega}<1$,
the following is true:
\begin{enumerate}
\item
The spectrum of $\eub{J}\eub{L}$
is symmetric with respect to the real
and imaginary axes.
\item The essential spectrum of $\eub{J}\eub{L}$
lies on the   imaginary axis and
is given by
\[
\sigma\sb{ess}(\eub{J}\eub{L})
= i\R\setminus(-i(1-\omega),i(1-\omega)).
\]
\item
$
\ker \eub{J}\eub{L}=\mathop{\rm Span}
\left\langle
\left[\begin{matrix}\p\sb x\phi\\0\end{matrix}\right],
\left[\begin{matrix}0\\\phi\end{matrix}\right]
\right\rangle,
$
with
$
\phi=\left[\begin{matrix}v\\u\end{matrix}\right]\in\R^2.
$
\item
The values $\lambda = \pm 2i\omega$ are eigenvalues of
$\eub{J}\eub{L}$.
The corresponding eigenvectors are
$
\left[
\begin{matrix}
\varphi\\\pm i\varphi
\end{matrix}\right]
$,
where
$\varphi=\left[\begin{matrix}u\\v\end{matrix}\right]$.
\end{enumerate}
\end{lemma}

\begin{remark}
More generally,
it turns out that $\lambda=\pm 2\omega i$
are $L\sp 2$-eigenvalues of
$\eub{JL}$
which corresponds to a linearization
at a solitary wave
of nonlinear Dirac equation \eqref{dirac}
with any nonlinearity $g(s)$ and in any dimension $n\ge 1$.
These are embedded eigenvalues
as long as $\abs{\omega}>m/3$, where $m:=g(0)$.
See {\rm \cite{dirac-vk-arxiv}}.
\end{remark}

\begin{proof}
Let us show that
$\sigma\sb p(\eub{JL})$
is symmetric with respect to
$\Re\lambda=0$
and
$\Im\lambda=0$.
Since
$\eub{J}\eub{L}
=\left[\begin{matrix}0&\eur{L}\sb{0}\\-\eur{L}\sb{1}&0\end{matrix}\right]$,
with
both $\eur{L}\sb\pm$ real-valued,
$\lambda\in\sigma(\eub{J}\eub{L})$
implies that
if $\varPsi\in L\sp{2}(\R,\C^4)$
satisfies $\eub{J}\eub{L}\varPsi=\lambda\varPsi$
(in the sense of distributions),
then
$\overline{\varPsi(x)}$
satisfies
$\eub{J}\eub{L}\bar\varPsi=\bar\lambda\bar\varPsi$.
At the same time,
the function
$\bm\varSigma\varPsi(x)$,
with
$\bm\varSigma=\left[\begin{matrix}
I\sb 2&0\\0&-I\sb 2
\end{matrix}\right],
$
satisfies
\[
\eub{J}\eub{L}\bm\varSigma\varPsi
=\eub{J}\bm\varSigma\eub{L}\varPsi
=-\bm\varSigma\eub{J}\eub{L}\varPsi
=-\lambda\bm\varSigma\varPsi.
\]
It follows that
both $\bar\lambda$ and $-\lambda$
are also eigenvalues of $\eub{JL}$,
hence
$\sigma\sb p(\eub{J}\eub{L})$
is symmetric with respect to the line $\Im\lambda=0$
and with respect to the point $\lambda=0$.

To find the essential spectrum of $\eub{J}\eub{L}$,
by the Weyl criterion,
we need to consider the limit of $\eub{J}\eub{L}$
as $x\to\pm\infty$,
substituting $v$, $u$ by
zeros and $g$ by $g(0)=m=1$;
then $\eub{J}\eub{L}-\lambda$
turns into $\eub{J}(\eub{D}-\omega)-\lambda$,
where
$\eub{D}=\left[\begin{matrix}\eur{D}&0\\0&\eur{D}\end{matrix}\right]$,
with $\eur{D}=
\left[
\begin{matrix}1&\p\sb x\\-\p\sb x&-1\end{matrix}\right]
$ defined in \eqref{eq:L-infinity-def}.
Substituting into
$(\eub{J}(\eub{D}-\omega)-\lambda)\varPsi=0$
the Ansatz
$\varPsi(x)=\varXi e^{i\xi x}$,
with
$\varXi\in\C^4$,
$\varXi\ne 0$,
we get:
\[
(\eub{J}(\eub{D}-\omega)-\lambda)\varXi
e^{i\xi x}
=
e^{i\xi x}
(\eub{J}(\eub{D}(\xi)-\omega)-\lambda)\varXi
=
e^{i\xi x}
\left[\begin{matrix}-\lambda&\eur{D}(\xi)-\omega\\-\eur{D}(\xi)+\omega&-\lambda\end{matrix}\right]
\varXi
=0,
\]
where
$\eub{D}(\xi)
=\left[\begin{matrix}\eur{D}(\xi)&0\\0&\eur{D}(\xi)\end{matrix}\right]$,
with
$\eur{D}(\xi)=
\left[
\begin{matrix}1&i\xi\\-i\xi&-1\end{matrix}\right]
$
being the symbol of
$\eur{D}$.
The essential spectrum of $\eub{J}\eub{L}$ is the range of values of
$\lambda$
which correspond to $\xi\in\R$.
To find the relation between $\lambda$ and $\xi$,
we need to compute the determinant
$\det\left(\eub{J}(\eub{D}(\xi)-\omega)-\lambda\right)$
and to equate it to zero.
In order to compute the determinant,
we notice that
\[
\big(\eub{J}(\eub{D}(\xi)-\omega)-\lambda\big)
\big(\eub{J}(\eub{D}(\xi)-\omega)+\lambda\big)
=
-(\eub{D}(\xi)-\omega)^2-\lambda^2
=-\xi^2-1-\omega^2-\lambda^2+2\omega\eub{D}(\xi),
\]
\[
\big(-\xi^2-1-\omega^2-\lambda^2+2\omega\eub{D}(\xi)\big)
\big(-\xi^2-1-\omega^2-\lambda^2-2\omega\eub{D}(\xi)\big)
=\big((1-\omega^2+\xi^2+\lambda^2)^2+4\omega^2\lambda^2\big){I}\sb 4,
\]
where ${I}\sb 4$ is the $4\times 4$ identity matrix.
Since
$\det\left(\eub{J}(\eub{D}(\xi)-\omega)-\lambda\right)$
should be even with respect to $\xi$ and $\omega$,
two above relations
allow us to conclude that
$
\det\left(\eub{J}(\eub{D}(\xi)-\omega)-\lambda\right)
=(1-\omega^2+\xi^2+\lambda^2)^2+4\omega^2\lambda^2.
$
The equation
\[
\det\left(\eub{J}(\eub{D}(\xi)-\omega)-\lambda\right)
=
(1-\omega^2+\xi^2+\lambda^2)^2
+4\omega^2\lambda^2
=0
\]
allows one to express $\lambda$ in terms of $\xi$
as
$
\lambda=\pm i(\omega\pm\sqrt{1+\xi^2}),
$
or, vice versa,
\[
\xi  = \pm\sqrt{ (\omega\pm i\lambda)^2 - 1}.
\]
This shows that the essential spectrum
is $i\R\backslash(-i(1-\omega),i(1-\omega))$.

The kernel of $\eub{J}\eub{L}$ is known
by Lemma~\ref{lemma-hm-2omega}.
Finally, again due to
Lemma~\ref{lemma-hm-2omega},
\[
\eub{J}\eub{L}
\left[
\begin{matrix}\varphi\\\pm i\varphi\end{matrix}\right]
=
\left[
\begin{matrix}0&\eur{L}\sb{0}\\-\eur{L}\sb{1}&0\end{matrix}\right]
\left[
\begin{matrix}\varphi\\\pm i\varphi\end{matrix}\right]
=
-2\omega
\left[
\begin{matrix}\pm i\varphi\\-\varphi\end{matrix}\right]
=
\mp 2\omega i
\left[
\begin{matrix}\varphi\\\pm i\varphi\end{matrix}\right].
\]
\end{proof}

\begin{definition}
Threshold points
are the values of $\lambda\in\C$
which correspond to $\xi=0$.
\end{definition}

On Figure~\ref{fig-sigma-c-xi},
one can see that the essential spectrum of
$\eub{JL}$
consists of two overlapping components,
which we distinguish by the symbols
$\flat$ and $\sharp$.
The $\flat$-component is
$i\R\backslash(\lambda\sp\flat\sb{d},\lambda\sp\flat\sb{u})$,
with the threshold points
\begin{equation}\label{tpf}
\lambda\sp\flat\sb{d}=-i-i\omega,
\qquad
\lambda\sp\flat\sb{u}=i-i\omega;
\end{equation}
the $\sharp$-component is
$i\R\backslash(\lambda\sp\sharp\sb{d},\lambda\sp\sharp\sb{u})$,
with the threshold points
\begin{equation}\label{tps}
\lambda\sp\sharp\sb{d}=-i+i\omega,
\qquad
\lambda\sp\sharp\sb{u}=i+i\omega.
\end{equation}
We use the subscripts ``$d$'' and ``$u$''
for the lower (``down'') and the upper
edges of each of the $\flat$, $\sharp$
components of the essential spectrum.

\begin{figure}[htbp]
\hskip -1cm
\includegraphics[width=17cm,height=12cm]{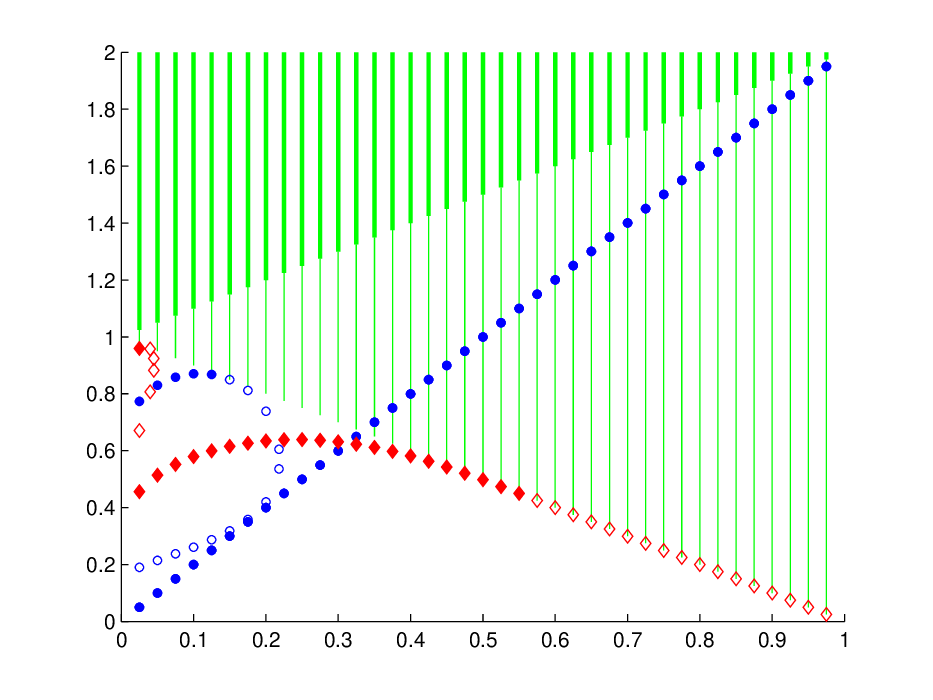}
\caption{\footnotesize 
$\sigma(\eub{J}\eub{L})$.
The zeros of the Evans function
located
in the upper half of the spectral gap (vertical axis)
as a function of $\omega$ (horizontal axis).
Eigenvalues ($\blackdiamond$, $\bullet$)
and the values of $\lambda$
corresponding to antibound states
($\diamond$, $\circ$).
The eigenvalue $2\omega i$
(the straight line of $\bullet$)
is embedded into the essential spectrum
for $\omega>1/3$.
}
\label{JL-spectrum-zoom}
\bigskip
\includegraphics[width=16cm,height=6cm]{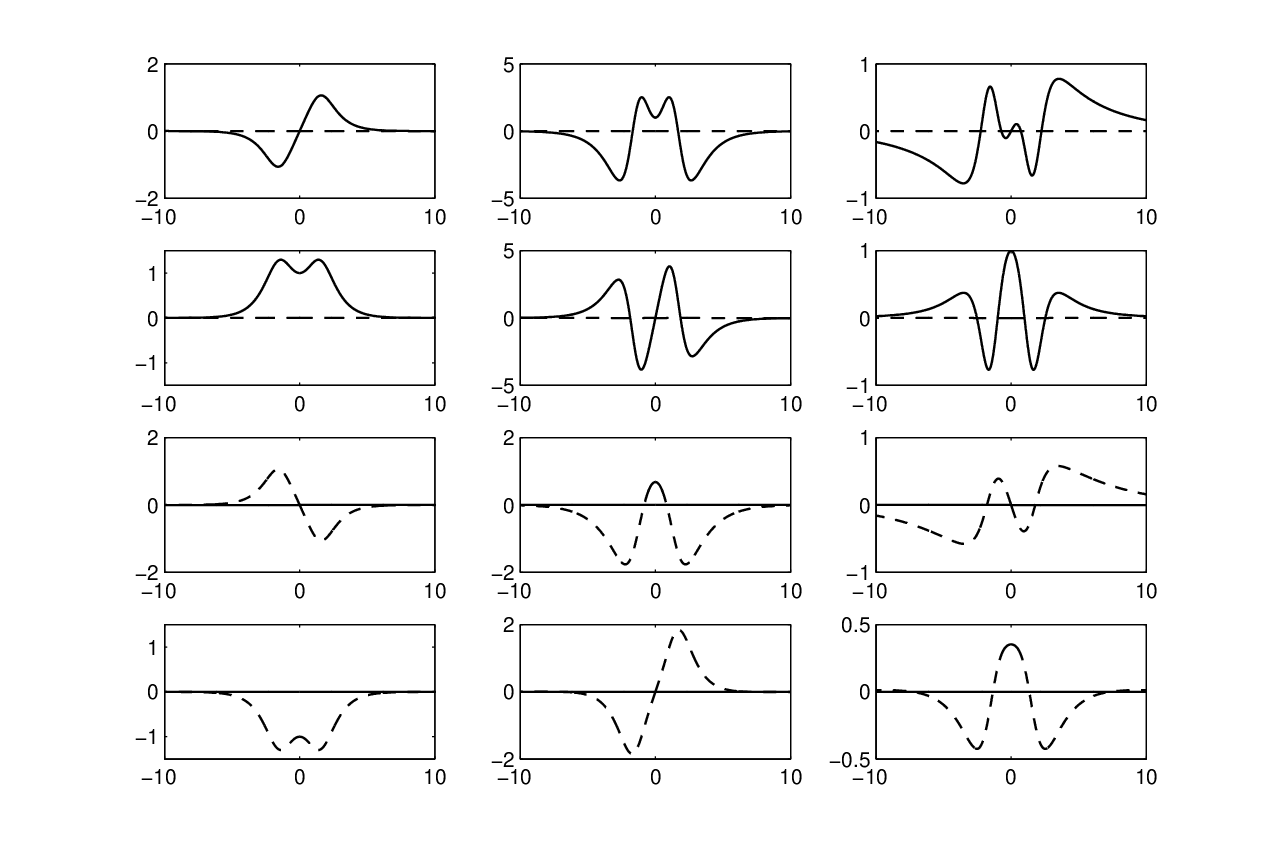}
\caption{\footnotesize
Components of the eigenfunctions
corresponding to the eigenvalues
$0.2000i$, $0.5792i$, and $0.8703i$
of $\eub{J}\eub{L}$,
located in the upper half of the spectral gap
for $\omega=0.1$.
All three eigenfunctions
have their
first two components real-valued (solid lines)
and
second two components imaginary (dashed lines).
}
\label{JL-eigenfunctions}
\end{figure}

The numerical computation
of the point spectrum of $\eub{J}\eub{L}$
inside $(0,i(1+\omega))$,
as a function of $\omega\in(0,1)$,
is plotted on Figure~\ref{JL-spectrum-zoom}.
The eigenfunctions
corresponding to three point eigenvalues 
at $\omega=0.1$
are plotted on
Figure~\ref{JL-eigenfunctions}.

\section{Jost solutions and Evans functions}
\label{sect-jost}

Looking for zeros of the Evans function is a way to test when an
equation has solutions with the correct asymptotics
at infinity.
The definition involves two
main steps.
The first step is to construct \emph{Jost solutions},
which are defined as solutions
with certain
decaying asymptotics either at plus infinity or at minus infinity.
The second
step is to match these two types of solutions.
In the presence of symmetry the
construction can be made simpler,
simplifying the numerical computations.
We describe this in detail below.

\subsection{Jost solutions for $\eub{J}\eub{L}$}

\begin{figure}[htbp]
\begin{center}
\includegraphics{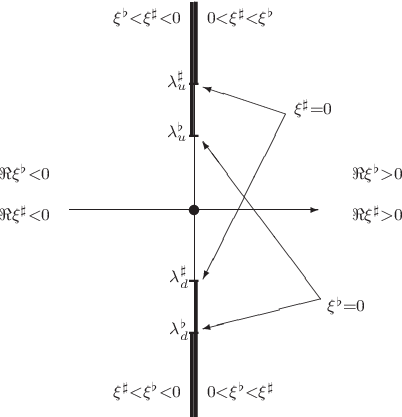}
\caption{\footnotesize 
Domains of analytic functions $\xi\sp\flat(\lambda)$,
$\xi\sp\sharp(\lambda)$,
and the essential spectrum of $\eub{J}\eub{L}$.
The thresholds are located at
$\lambda\sp\flat\sb{d}=-i-i\omega$,
$\lambda\sp\sharp\sb{d}=-i+i\omega$,
$\lambda\sp\flat\sb{u}=i-i\omega$,
and
$\lambda\sp\sharp\sb{u}=i+i\omega$.
The analytic function
$\xi\sp\flat(\lambda)$
is defined on $\C$ with the cuts from $\lambda\sp\flat\sb{u}$ to $+i\infty$
and from $\lambda\sp\flat\sb{d}$ to $-i\infty$
(thick lines on the left side of the imaginary axis).
$\Im\xi\sp\flat<0$, $\Im\xi\sp\sharp<0$
for all $\lambda\in\C$.
The analytic function
$\xi\sp\sharp(\lambda)$
is defined on $\C$ with the cuts from $\lambda\sp\sharp\sb{u}$ to $+i\infty$
and from $\lambda\sp\sharp\sb{d}$ to $-i\infty$
(thick lines on the right side of the imaginary axis).
The boundary traces of
both $\xi\sp\sharp$ and $\xi\sp\flat$
at the double-covered
part of the essential spectrum
(above $\lambda\sp\sharp\sb{u}$ and below $\lambda\sp\flat\sb{d}$)
are real-valued.
}
\label{fig-sigma-c-xi}
\end{center}
\end{figure}

Jost solutions $Y(x,\lambda)$
of $\eub{J}\eub{L}$
are defined as solutions to
$(\eub{J}\eub{L}-\lambda)\varPsi=0$
which have the same asymptotics at $+\infty$
or at $-\infty$
as the solutions to
$(\eub{J}(\eub{D}-\omega)-\lambda)\varPsi=0$,
where
\[
\eub{D}=
\left[\begin{matrix}
\eur{D}&0\\0&\eur{D}
\end{matrix}\right],
\qquad
\eur{D}=
-i\alpha\p\sb x+\beta
=\left[\begin{matrix}1&\p\sb x\\-\p\sb x&-1\end{matrix}\right].
\]
For a given $\lambda\in\C$
away from the threshold points $\pm(1\pm\omega)i$,
solutions to
$(\eub{J}(\eub{D}-\omega)-\lambda)\varPsi=0$
have the form
\[
\varPsi(x,\lambda)
=\left[\begin{matrix}R\\S\end{matrix}\right]e^{i\xi x},
\]
where
$\left[\begin{matrix}R\\S\end{matrix}\right]\in\C^4$
and $\xi\in\C$ is a solution to
$\det\left(\eub{J}(\eub{L}(\xi)-\omega)-\lambda\right)=0$.
Let $\lambda=a+ib$,
with $a\ge 0$ and $b\ge 0$.
(Because of the symmetry of $\sigma(\eub{JL})$
with respect to the lines $\Re\lambda=0$ and $\Im\lambda=0$,
we only need to consider the spectrum
in the closure of the first quadrant of $\C$.)
Then define
\[
\xi\sp\flat
=\sqrt{((b-ia)+\omega)^2-1},
\qquad
\qquad
\xi\sp\sharp
=\sqrt{((b-ia)-\omega)^2-1},
\]
where for the square root we choose the branch that has negative
imaginary part (so that
both
$e^{-i\xi\sp\flat x}$
and
$e^{-i\xi\sp\sharp x}$
decay as $x\to+\infty$).
Then $\xi\sp\flat$
is defined for $\lambda\in\C$
with the cuts from $\lambda\sp\flat\sb{u}=i(1-\omega)$ to $+i\infty$
and from $\lambda\sp\flat\sb{d}=-i(1+\omega)$ to $-i\infty$,
while $\xi\sp\sharp$
is defined for $\lambda\in\C$
with the cuts from $\lambda\sp\sharp\sb{u}=i(1+\omega)$ to $+i\infty$
and from $\lambda\sp\sharp\sb{d}=-i(1-\omega)$ to $-i\infty$.
See Figure~\ref{fig-sigma-c-xi}.
Altogether the four solutions to the equation
$\det\left(\eub{J}(\eub{D}(\xi)-\omega)-\lambda\right)=0$
are $\pm\xi\sp\flat(\lambda)$
and $\pm\xi\sp\sharp(\lambda)$.

The four solutions to
$(\eub{J}(\eub{D}-\omega)-\lambda)\varPsi=0$
corresponding
to $\lambda$
away from the thresholds
\eqref{tpf}, \eqref{tps}
are given by
\begin{equation}
\varXi\sp\flat\sb{\pm}(\lambda)e^{\pm\xi\sp\flat(\lambda)x},
\qquad
\varXi\sp\sharp\sb{\pm}(\lambda)e^{\pm\xi\sp\sharp(\lambda)x},
\end{equation}
where
\begin{eqnarray}
\varXi\sp\flat\sb{-}(\lambda)
=\left[\begin{matrix}
-i\xi\sp\flat(\lambda)
\\
-i\lambda-1+\omega
\\
\xi\sp\flat(\lambda)
\\
\lambda-i(1-\omega)
\end{matrix}\right],
\qquad
\varXi\sp\sharp\sb{-}(\lambda)
=\left[\begin{matrix}
-i\xi\sp\sharp(\lambda)
\\
i\lambda-1+\omega
\\
-\xi\sp\sharp(\lambda)
\\
\lambda+i(1-\omega)
\end{matrix}\right],
\label{mpmm}
\\ \nonumber \phantom{\int}\\
\qquad
\varXi\sp\flat\sb{+}(\lambda)
=\left[\begin{matrix}
i\xi\sp\flat(\lambda)
\\
-i\lambda-1+\omega
\\
-\xi\sp\flat(\lambda)
\\
\lambda-i(1-\omega)
\end{matrix}\right],
\qquad
\varXi\sp\sharp\sb{+}(\lambda)
=\left[\begin{matrix}
i\xi\sp\sharp(\lambda)
\\
i\lambda-1+\omega
\\
\xi\sp\sharp(\lambda)
\\
\lambda+i(1-\omega)
\end{matrix}\right].
\label{pppm}
\end{eqnarray}

We will only be considering the Jost solutions
which have prescribed asymptotics
at $x\to +\infty$.

\begin{lemma}\label{lemma-jost}
For each $\lambda\in\C$,
$\lambda\notin
\{
\lambda\sp\flat\sb{d},
\lambda\sp\sharp\sb{d},
\lambda\sp\flat\sb{u},
\lambda\sp\sharp\sb{u}\}$,
there are Jost solutions to $(\eub{JL}-\lambda)\varPsi=0$
with the asymptotics
$Y\sp\sharp\sb{\pm}(x,\lambda)
\sim\varXi\sp\sharp\sb{\pm}(\lambda)e^{\pm i\xi\sp\sharp(\lambda)x}$
and
$Y\sp\flat\sb{\pm}(x,\lambda)
\sim\varXi\sp\flat\sb{\pm}(\lambda)e^{\pm i\xi\sp\flat(\lambda)x}$,
$x\to+\infty$.
More precisely,
\[
\abs{
Y\sp\sharp\sb{\pm}(x,\lambda)e^{\mp i\xi\sp\sharp(\lambda)x}
-\varXi\sp\sharp\sb{\pm}(\lambda)}
=o(1),
\qquad x\to +\infty;
\]
\[
\abs{
Y\sp\flat\sb{\pm}(x,\lambda)e^{\mp i\xi\sp\flat(\lambda)x}
-\varXi\sp\flat\sb{\pm}(\lambda)}
=o(1),
\qquad x\to +\infty.
\]
\end{lemma}

\begin{proof}
The proof follows from the Duhamel representation
for the solution
to $(\eub{JL}-\lambda)\varPsi=0$
and from the exponential
spatial
decay of the
solitary waves
$\phi(x)$
corresponding to $\omega\in(0,1)$;
see Remark~\ref{remark-ed}.
\end{proof}

\begin{remark}
At the threshold points
$\lambda\sp\sharp\sb{d}=-i+i\omega$
and
$\lambda\sp\sharp\sb{u}=i+i\omega$
(respectively,
$\lambda\sp\flat\sb{d}=-i-i\omega$
and
$\lambda\sp\flat\sb{u}=i-i\omega$),
where
$\xi\sp\sharp(\lambda)=0$
(respectively, $\xi\sp\flat(\lambda)=0$),
one has
$\varXi\sp\sharp\sb{+}(\lambda)=\varXi\sp\sharp\sb{-}(\lambda)$
(respectively,
$\varXi\sp\flat\sb{+}(\lambda)=\varXi\sp\flat\sb{-}(\lambda)$).
For such $\lambda$,
there are only three Jost solutions
as in Lemma~\ref{lemma-jost},
and one more Jost solution which is linearly growing
as $x\to+\infty$.
\end{remark}

\subsection{Evans functions for $\eub{JL}$}

Normally, Evans function describes a matching between Jost solutions
decaying to the left and Jost solutions decaying to the right.
However, presence of symmetries allows us to streamline calculation in
the present case.
Denote by $X\sp{\blackdiamond}$ the ``even'' subspace of functions from
$C^1(\R,\C^4)$ with even first and third components and with odd second
and fourth components.  Similarly, denote by $X\sp{\bullet}$ the
``odd''
subspace in $C^1(\R,\C^4)$ with odd first and third components and with
even second and fourth components.  Then
$C^1(\R,\C^4)=X\sp{\blackdiamond}\oplus X\sp{\bullet}$.
Noticing that
$\eub{J}\eub{L}$
acts invariantly in $X\sp{\blackdiamond}$ and in $X\sp{\bullet}$,
we conclude that all eigenvalues of $\eub{J}\eub{L}$
always have a corresponding eigenfunction either in
$X\sp{\blackdiamond}$ or in $X\sp{\bullet}$ (or in both subspaces).
To find
eigenvalues of $\eub{J}\eub{L}$ corresponding to functions from
$X\sp{\blackdiamond}$, we proceed as follows:

\begin{itemize}
\item
For $\lambda\in\C$,
construct solutions
$\varPsi\sb{j}$,
$1\le j\le 4$,
to the equation $\eub{J}\eub{L}\varPsi=\lambda\varPsi$
with the following initial data
at $x=0$:
\begin{equation}\label{vp1234}
\varPsi\sb 1\att{x=0}
={\tiny\left[\begin{matrix}1\\0\\0\\0\end{matrix}\right]},
\qquad
\varPsi\sb 2\att{x=0}
={\tiny\left[\begin{matrix}0\\1\\0\\0\end{matrix}\right]},
\qquad
\varPsi\sb 3\att{x=0}
={\tiny\left[\begin{matrix}0\\0\\1\\0\end{matrix}\right]},
\qquad
\varPsi\sb 4\att{x=0}
={\tiny\left[\begin{matrix}0\\0\\0\\1\end{matrix}\right]}.
\end{equation}
Then
$\varPsi\sb{1}$, $\varPsi\sb{3}\in X\sp{\blackdiamond}$,
while
$\varPsi\sb{2}$, $\varPsi\sb{4}\in X\sp{\bullet}$.
\item
Take the Jost solutions
$Y\sp\flat\sb{-}(x,\lambda)$ and $Y\sp\sharp\sb{-}(x,\lambda)$
as in Lemma~\ref{lemma-jost},
which decay for $x\to+\infty$.
\item Define
the Evans function
\begin{equation}\label{efmm-13}
E\sb{--}\sp{\blackdiamond}(\lambda)
=\det\left[\varPsi\sb{1}(x,\lambda),\,\varPsi\sb{3}(x,\lambda),\,
Y\sp\flat\sb{-}(x,\lambda),\,Y\sp\sharp\sb{-}(x,\lambda)\right].
\end{equation}
This is a Wronskian-type
function which does not depend on $x$ and could be evaluated at
$x=R\gg 1$, where the asymptotics of $Y\sp\flat\sb{-}$ and
$Y\sp\sharp\sb{-}$ are known from Lemma~\ref{lemma-jost}.
Vanishing of $E\sb{--}\sp{\blackdiamond}(\lambda)$
at particular $\lambda\in\C$
means that a certain linear combination
of $\varPsi\sb 1(x,\lambda)$ and $\varPsi\sb 3(x,\lambda)$
has the asymptotics of
the linear combination
of
$Y\sp\flat\sb{-}(x,\lambda)$ and $Y\sp\sharp\sb{-}(x,\lambda)$
as $x\to+\infty$,
which decays at $+\infty$
(according to our choice of $\xi\sp\flat(\lambda)$ and $\xi\sp\sharp(\lambda)$.
By the symmetry of $\varPsi$
(its first and third components are even
while its second and fourth components are odd),
this same linear combination
also decays as $x\to-\infty$.
Therefore,
vanishing of $E\sb{--}\sp{\blackdiamond}(\lambda)$
at some $\lambda\in\C$
implies that
there is an eigenfunction corresponding to this particular
value of $\lambda$.
\item
Similarly,
define
\begin{equation}\label{efmm-24}
E\sb{--}\sp{\bullet}(\lambda)
=
\det\left[\varPsi\sb{2}(x,\lambda),\,\varPsi\sb{4}(x,\lambda),\,
Y\sp\flat\sb{-}(x,\lambda),\,Y\sp\sharp\sb{-}(x,\lambda)\right].
\end{equation}
The condition
$E\sp{\bullet}(\lambda)=0$
means that a certain linear combination
of
$\varPsi\sb{2}$, $\varPsi\sb{4}\in X\sp{\bullet}$
has the same asymptotics when $x\to+\infty$
as a solution
of $(\eub{J}(\eub{D}-\omega)-\lambda)\varPsi=0$
which decays for $x\to +\infty$.
\end{itemize}

Let us summarize the above
in a convenient form:

\begin{lemma}
$\lambda\in\spec\sb{p}(\eub{J}\eub{L})
$
if and only if
$E\sb{--}\sp{\blackdiamond}(\lambda)E\sb{--}\sp{\bullet}(\lambda)=0$.
Furthermore, $E\sb{--}\sp{\blackdiamond}(\lambda)=0$
(respectively,
$E\sb{--}\sp{\bullet}(\lambda)=0$)
if the corresponding wave function
belongs to $L\sp 2(\R,\C^4)\cap X\sp{\blackdiamond}$
(respectively, $L\sp 2(\R,\C^4)\cap X\sp{\bullet}$).
\end{lemma}

When searching numerically
for zeros
of the Evans functions
in the spectral gap on the imaginary axis,
we benefit from the following observation.

\begin{lemma}
For $\lambda\in i(-(1-\omega),(1-\omega))$,
the real part of the functions
$
E\sb{--}\sp{\blackdiamond}(\lambda),
$
$
E\sb{--}\sp{\bullet}(\lambda)
$
equals zero.
\end{lemma}

\begin{proof}
For $\lambda$ with $\Re\lambda=0$,
one immediately concludes
from $(\eub{J}\eub{L}-\lambda)\varPsi\sb j=0$
that for all $x\in\R$
the components $(\varPsi\sb 1(x))\sb j$
and $(\varPsi\sb 2(x))\sb j$
are real for $j=1,\,2$
and imaginary for $j=3,\,4$,
and that $(\varPsi\sb 3(x))\sb j$
and $(\varPsi\sb 4(x))\sb j$
are imaginary for $j=1,\,2$ and real for $j=3,\,4$.
On the other hand, when $\lambda\in(-i(1-\omega),i(1-\omega))$
and both $\xi\sp\flat$ and $\xi\sp\sharp$ are imaginary,
the first two components of $\varXi\sp\flat\sb{-}$ and $\varXi\sp\sharp\sb{-}$
from \eqref{mpmm}
are real and the second two are imaginary.
It follows that
\[
\det\left[\varPsi\sb 1(x,\lambda),\varPsi\sb 3(x,\lambda),
\varXi\sp\flat\sb{-}(\lambda),\varXi\sp\sharp\sb{-}(\lambda)\right]
\in i\R,
\qquad
\det\left[\varPsi\sb 2(x,\lambda),\varPsi\sb 4(x,\lambda),
\varXi\sp\flat\sb{-}(\lambda),\varXi\sp\sharp\sb{-}(\lambda)\right]
\in i\R,
\]
for any $x\in\R$.
Since
$\xi\sp\flat$ and $\xi\sp\sharp$ are purely imaginary,
$e^{-i\xi\sp\flat x}$
and $e^{-i\xi\sp\sharp x}$
are real;
we conclude from
Lemma~\ref{lemma-jost}
that
\[
E\sb{--}\sp{\blackdiamond}(\lambda)
=
\det\left[\varPsi\sb 1(x,\lambda),\varPsi\sb 3(x,\lambda),
Y\sp\flat\sb{-}(\lambda),Y\sp\sharp\sb{-}(\lambda)\right]
\]
and
\[
E\sb{--}\sp{\bullet}(\lambda)
=
\det\left[\varPsi\sb 2(x,\lambda),\varPsi\sb 4(x,\lambda),
Y\sp\flat\sb{-}(\lambda),Y\sp\sharp\sb{-}(\lambda)\right]
\]
are purely imaginary.
\end{proof}

\subsection{Jost solutions
and Evans functions
for $\eur{L}\sb{0}$ and $\eur{L}\sb{1}$}

The construction of the Jost solutions and
Evans function for the operator $\eur{L}\sb{1}$
(and, respectively, $\eur{L}\sb{0}$) 
is similar
to the construction for $\eub{J}\eub{L}$.
At $x\to\pm\infty$,
$\eur{L}\sb{1}$ coincides with $\eur{D}-\omega$.
The equation
\[
(\eur{D}-\omega-\lambda)\Psi(x)=0
\]
has two linearly independent solutions
\[
\Xi\sb\pm(\lambda)e^{\pm i\xi(\lambda) x},
\]
where $\Xi\sb\pm(\lambda)\in\C^2$
are given by
\[
\Xi\sb\pm(\lambda)
=
\left[
\begin{array}{c}
1+ \omega+\lambda\\
\mp i\xi(\lambda)
\end{array}
\right],
\qquad \mbox{where}\quad \xi = \sqrt{(\omega+\lambda)^2-1}.
\]
The function $\xi(\lambda)$ is defined for $\lambda\in\C$ with
branch cuts from $\lambda= 1-\omega$ to $+\infty$ and from
$\lambda=-1-\omega$ to $-\infty$.
These branch cuts correspond to the
essential spectrum of the operator $\eur{L}\sb{1}$
(similarly, of $\eur{L}\sb{0}$).
The square root
denotes the branch
with the negative imaginary part
when the argument is negative,
so that
for $\lambda$ from the spectral gap
$(-1-\omega,1-\omega)$,
the function $\Xi\sb{-}(\lambda)e^{-i\xi(\lambda)x}$
is decaying as
$x\to+\infty$.
The Jost solutions
${\rm Y}\sb\pm(x,\lambda)$
for $\eur{L}\sb{1}$
are solutions to $(\eur{L}\sb{1}-\lambda){\rm Y}(x,\lambda)=0$
with the asymptotic behavior
\begin{equation}
\label{eq:Jost_Lpm}
{\rm Y}\sb\pm(x,\lambda)\sim\Xi\sb\pm(\lambda)e^{\pm i\xi(\lambda) x},
\qquad
x\to+\infty.
\end{equation}
There are two subspaces of $C^1(\R, \C^2)$,
${\rm X}\sp{\blackdiamond}$ (spinors
with the even first component and odd second component)
and ${\rm X}\sp{\bullet}$
(spinors with the odd first component and even second component)
such that
${\rm X}\sp{\blackdiamond}\oplus {\rm X}\sp{\bullet}=C^1(\R,\C^2)$,
which are invariant with respect to
$\eur{L}\sb{1}$
(also with respect to $\eur{L}\sb{0}$).
We define two solutions,
$\Psi_1(x,\lambda)$ and $\Psi_2(x,\lambda)$,
to the equation $(\eur{L}\sb{1}-\lambda)\Psi=0$,
with the initial data
\begin{equation*}
\Psi\sb 1\att{x=0}
=\left[\begin{matrix}1\\0\end{matrix}\right],
\qquad
\Psi\sb 2\att{x=0}
=\left[\begin{matrix}0\\1\end{matrix}\right]. 
\end{equation*}
Then we define the Evans functions
of $\eur{L}\sb 1$
by
\begin{equation}\label{def-em}
{\rm E}\sb{\pm}\sp{\blackdiamond}(\lambda)
=\det\left[\Psi_1(x,\lambda),\,{\rm Y}\sb{\pm}(x,\lambda)\right],
\qquad
{\rm E}\sb{\pm}\sp{\bullet}(\lambda)
=\det\left[\Psi_2(x,\lambda),\,{\rm Y}\sb{\pm}(x,\lambda)\right],
\end{equation}
and look for their zeros.

If ${\rm E}\sb{-}\sp\blackdiamond(\lambda)$
vanishes at some $\lambda\in\C$,
then it means that
$\Psi\sb 1(x,\lambda)$
as $x\to+\infty$ has the asymptotics of the
decaying Jost solution.
(By the symmetry,
as $x\to-\infty$, $\Psi\sb 1(x,\lambda)$
also has the asymptotics of the
Jost solution decaying to $-\infty$.)
We can summarize this as follows.

\begin{lemma}
The inclusion $\lambda\in\sigma\sb{p}(\eur{L}\sb{1})$
takes place if and only if
${\rm E}\sb{-}\sp{\blackdiamond}(\lambda){\rm E}\sb{-}\sp{\bullet}(\lambda)=0$.
\end{lemma}

In the same way
one defines the Evans functions
${\rm E}\sp{\blackdiamond}\sb{-}(\lambda)$
and ${\rm E}\sp{\bullet}\sb{-}(\lambda)$
for $\eur{L}\sb{0}$.
The zeros
of the Evans functions
${\rm E}\sp{\blackdiamond}\sb\pm(\lambda)$
and ${\rm E}\sp{\bullet}\sb\pm(\lambda)$
are plotted
on
Figure~\ref{Hminus-spectrum}
(for $\eur{L}\sb{0}$)
and
Figure~\ref{Hplus-spectrum}
(for $\eur{L}\sb{1}$).
The meaning of zeros of
${\rm E}\sb{+}\sp\blackdiamond$
and
${\rm E}\sb{+}\sp\bullet$
is discussed in Section~\ref{sec:antibound}.

\subsection{Antibound states for $\eur{L}\sb{0}$ and $\eur{L}\sb{1}$}
\label{sec:antibound}

The Evans function is defined using two pieces of data: a solution with
the given initial data and the Jost solution that corresponds to one
value of $\xi$, see equation \eqref{eq:Jost_Lpm}.  However, we can
view $\xi(\lambda)$ as defined on a Riemann surface with two sheets
(corresponding to $+\sqrt{\cdot}$ and $-\sqrt{\cdot}$) and two
singularity points at $\pm1-\omega$.  The two sheets are glued across
the cuts $(-\infty, -1-\omega]$ and $[1-\omega, \infty)$.  The two
eigenvectors of $\eur{D}-\omega$
(this operator
coincides with $\eur{L}\sb 0$ and $\eur{L}\sb 1$ at $x\to\pm\infty$)
can be thought of as the same
eigenvector that changes according to which sheet $\lambda$ is on.
Continuing in this
vein,
we consider
two previously defined Jost solutions
${\rm Y}\sb\pm(x,\lambda)$,
$\lambda\in\C\backslash\big((-\infty,-1-\omega)\cup(1-\omega,+\infty)\big)$,
as one Jost solution defined on the Riemann surface,
which is glued of two copies of
$\C\backslash\big((-\infty,-1-\omega)\cup(1-\omega,+\infty)\big)$.
We use this Jost solution
to define the Evans function on this Riemann surface.
Thus, for $\lambda$ on the first sheet of the Riemann surface,
the Evans functions
${\rm E}\sp{\blackdiamond}(\lambda)$, ${\rm E}\sp{\bullet}(\lambda)$
are represented by
${\rm E}\sp{\blackdiamond}\sb{-}(\lambda)$,
${\rm E}\sp{\bullet}\sb{-}(\lambda)$
from \eqref{def-em},
while on the second sheet
they are represented by
${\rm E}\sp{\blackdiamond}\sb{+}(\lambda)$,
${\rm E}\sp{\bullet}\sb{+}(\lambda)$.
When a zero of the Evans function disappears at the end of the
spectral gap, it does not ``dissolve''
in the essential spectrum,
but,
rather, it goes back into the gap, albeit on a different sheet of
the Riemann surface on which the Evans function is defined.  Such an
``unphysical'' zero of the Evans function is known in the literature
as a ``resonance'' or an ``antibound state''.  Since the ``resonance''
is also a name used specifically for a bounded solution at the
threshold of the essential spectrum (at the threshold, the two
notions coincide), we will be using the ``antibound state'' as the
name of choice.  It is ``antibound'' since the solution is purely
exponentially increasing as
$x\to\pm\infty$, consisting
solely of
${\rm Y}\sb{+}(x,\lambda)$
as $x\to+\infty$.

On Figure~\ref{Hplus-spectrum} the antibound states
of $\eur{L}\sb{1}$
are indicated by
transparent symbols ($\diamond$
is for the states with even eigenfunctions
and $\circ$ is for the states with odd
eigenfunctions).
Sometimes antibound states pass from the unphysical sheet
onto the physical one at the threshold point $\lambda=1-\omega$.
Note that
the curve of transparent circles on the right has a maximum.
This is the value
of $\omega$ (on the vertical axis) at which two zeros of the Evans function
living
\emph{off the real axis $\Im\lambda=0$} on the unphysical sheet collide and
create two zeros on the real axis.  The self-adjointness of the
operator $\eur{L}\sb{1}$ forbids such a behaviour on the physical sheet, but
it is possible on the unphysical one.

Antibound states
for the operator $\eur{L}\sb{0}$
are plotted on Figure~\ref{Hminus-spectrum}.

\subsection{Antibound states for $\eub{JL}$}

The Riemann surface on which the Evans function of the operator
$\eub{JL}$ is defined is similar but more complicated.  Indeed, the
two limiting frequencies $\xi\sp\flat$ and $\xi\sp\sharp$ are defined on a
two-sheeted surface each, but the surfaces are different.  The Evans
function is then defined on four sheets.  We will denote them by
$(+,+)$, $(+,-)$, $(-,+)$, $(-,-)$, depending on the sign in front of
$(\xi\sp\flat,\xi\sp\sharp)$.
The sheet $(-,-)$ is the physical one,
in the sense that the zeros of
the Evans function on this sheet are the eigenvalues of
the operator $\eub{JL}$.

The sheets are glued in the following manner.

\begin{verse}
\it Across the cuts
$(\lambda\sp\sharp\sb{u}, i\infty)$ and $(\lambda\sp\flat\sb{d}, -i\infty)$,
the sheet
$(+,+)$ is glued to $(-,-)$,
while the sheet $(+,-)$ is glued to $(-,+)$
(that is, both signs change to their opposites).
\end{verse}

\begin{verse}
\it Across the cut
$(\lambda\sp\flat\sb{u},\lambda\sp\sharp\sb{u})$,
the gluing is $(+,\cdot) \leftrightarrow
(-,\cdot)$ (only the sign of $\xi\sp\flat$ changes),
while across the cut
$(\lambda\sp\flat\sb{d},\lambda\sp\sharp\sb{d})$
the sign of $\xi\sp\sharp$
changes: $(\cdot,+)
\leftrightarrow (\cdot,-)$.
\end{verse}

The four branches of the Evans function $E\sp{\blackdiamond}$
on these sheets
could be written as follows:
\[
E\sp{\blackdiamond}\sb{--}(\lambda)=\det\left[
\varPsi\sb 1(x,\lambda),\varPsi\sb 3(x,\lambda),
Y\sp\flat\sb{-}(x,\lambda),Y\sp\sharp\sb{-}(x)
\right],
\]
\[
E\sp{\blackdiamond}\sb{+-}(\lambda)=\det\left[
\varPsi\sb 1(x,\lambda),\varPsi\sb 3(x,\lambda),
Y\sp\flat\sb{+}(x),Y\sp\sharp\sb{-}(x)
\right],
\]
\[
E\sp{\blackdiamond}\sb{-+}(\lambda)=\det\left[
\varPsi\sb 1(x,\lambda),\varPsi\sb 3(x,\lambda),
Y\sp\flat\sb{-}(x,\lambda),Y\sp\sharp\sb{+}(x)
\right],
\]
\[
E\sp{\blackdiamond}\sb{++}(\lambda)=\det\left[
\varPsi\sb 1(x,\lambda),\varPsi\sb 3(x,\lambda),
Y\sp\flat\sb{+}(x),Y\sp\sharp\sb{+}(x)
\right].
\]
Similarly
one defines the four branches of the Evans function
$E\sp{\bullet}$.

\begin{figure}[htbp]
\hskip -1.5cm
\includegraphics[width=18.5cm,height=18.5cm]{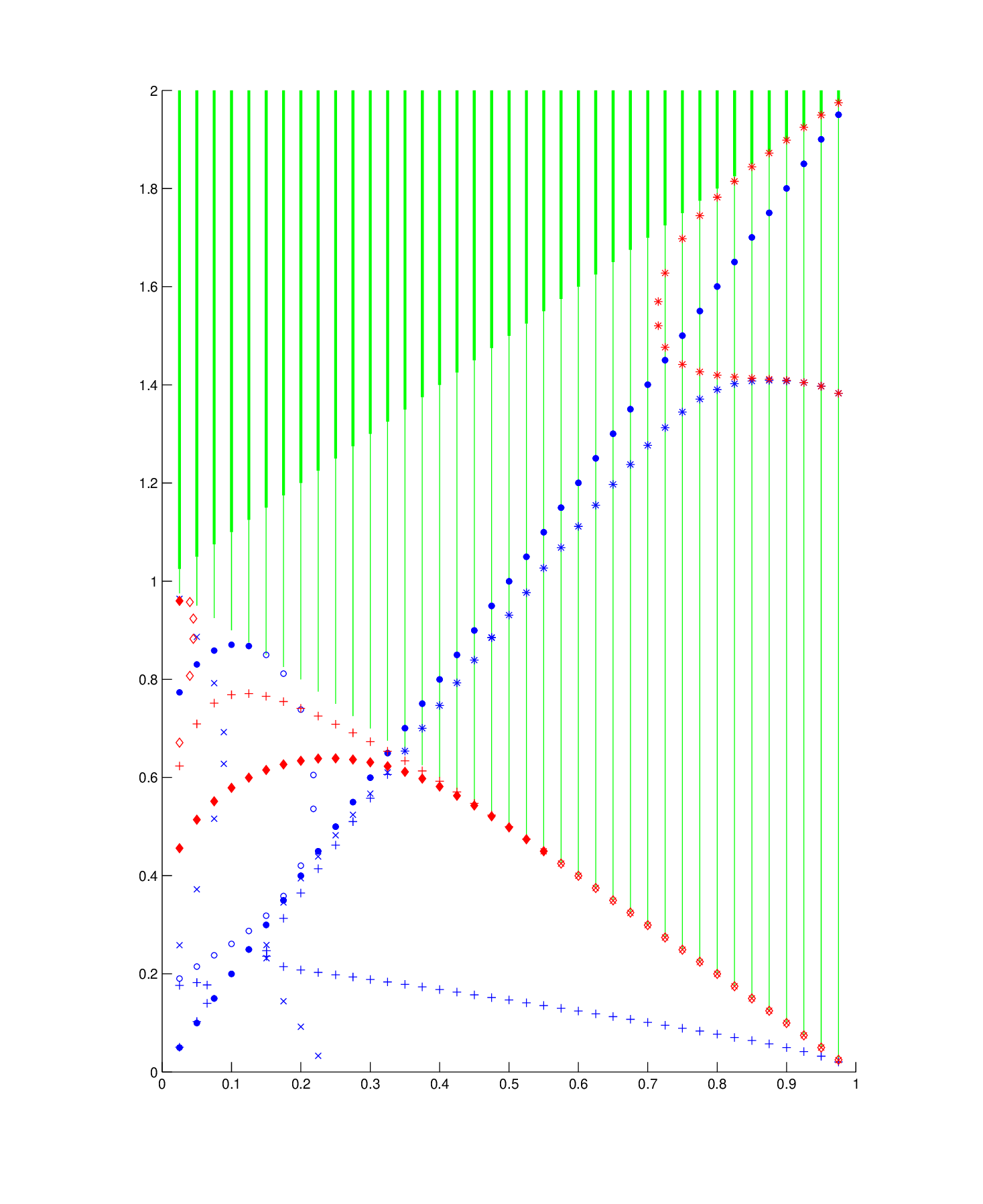}

\caption{\footnotesize 
$\sigma(\eub{J}\eub{L})$.
The zeros of the Evans function
on the upper half of the imaginary axis
(vertical)
as a function of $\omega$ (horizontal axis).
Eigenvalues ($\blackdiamond$ for even eigenfunctions,
$\bullet$ for odd)
and the values of $\lambda$
corresponding to antibound states
($\diamond$ for even, $\circ$ for odd).
The symbols ''$+$'' and ``$\times$''
denote zeros of the Evans functions
which correspond to
to the Jost solutions
on the other unphysical sheets
(sheets $(-,+)$ and $(+,+)$)
of the Riemann surface;
see Section~\ref{sec:antibound}.
The star symbols found inside the essential spectrum are actually made
up of coinciding symbols ``$+$'' and ``$\times$'';
see Lemma~\ref{lem:shooting_stars}.
}
\label{JL-spectrum-long}
\end{figure}

In Figure~\ref{JL-spectrum-long} we trace the zeros of the Evans function
on the $(-,-)$ sheet (eigenvalues, solid symbols) as well as the
zeros on the $(+,-)$ sheet (``antibound states'', transparent symbols).
The zeros can change
between the two sheets by hitting the (square root type) singularity
at $\lambda\sp\flat\sb{u}$.
Note that when a curve has infinite derivative
(with respect to $\omega$ on the $x$-axis)
it signals that the zeros of
Evans function are leaving the imaginary axis
into the complex plane away from $\Re\lambda=0$.
This behaviour can be seen for
zeros on the $(+,-)$ sheet, but we have not observed it for the
eigenvalues,
which are
the zeros on the $(-,-)$ sheet.
This suggests that the eigenvalues stay on the imaginary
axis for all values of $\omega$.

The zeros lying on the other two sheets are unlikely to
sneak onto the ``physical'' $(-,-)$ sheet
to become eigenvalues for the following reason.
To pass onto this sheet,
they would either have to leave the imaginary axis and circle
around or to go inside the essential spectrum and hit the singularity at
the embedded threshold at $\lambda\sp\sharp\sb{u}$.
We have not observed such a hypothetical behaviour.

For completeness,
we also plot
on Figure~\ref{JL-spectrum-long}
the zeros
of Evans functions
on the $(-,+)$ sheet
and on the $(+,+)$ sheet.
Note that
between the thresholds $\lambda\sp\flat\sb{u}$ and $\lambda\sp\sharp\sb{u}$,
these zeros
(marked on Figure~\ref{JL-spectrum-long} with ``$+$'' and ``$\times$'')
meet.
Indeed, there is the following simple observation.

\begin{lemma}
\label{lem:shooting_stars}
For
$\lambda
\in(\lambda\sp\flat\sb{d},\lambda\sp\sharp\sb{d})
\cup(\lambda\sp\flat\sb{u},\lambda\sp\sharp\sb{u})
$,
\[
E\sb{-+}\sp{\blackdiamond}(\lambda)=\overline{E\sb{++}\sp{\blackdiamond}(\lambda)},
\qquad
E\sb{-+}\sp{\bullet}(\lambda)=\overline{E\sb{++}\sp{\bullet}(\lambda)}.
\]
\end{lemma}

\begin{proof}
First, we notice that for $\lambda\in i\R$,
if $\varPsi$ is a solution to
\begin{equation}\label{asdf}
\eub{JL}\varPsi
=
\left[
\begin{matrix}
0&\eur{L}\sb 0
\\
-\eur{L}\sb 1&0
\end{matrix}
\right]
\varPsi
=\lambda\varPsi,
\end{equation}
then so is
$\bm\varSigma\overline{\varPsi}$,
where
$\bm\varSigma=\left[\begin{matrix}{I}\sb 2&0\\0&-{I}\sb 2\end{matrix}\right]$.
  From \eqref{vp1234},
we conclude that
for $\lambda\in i\R$,
\begin{eqnarray}\label{vp13}
\overline{\varPsi\sb 1(x,\lambda)}=\bm\varSigma\varPsi\sb 1(x,\lambda),
\qquad
\overline{\varPsi\sb 2(x,\lambda)}=\bm\varSigma\varPsi\sb 2(x,\lambda),
\noindent
\\
\overline{\varPsi\sb 3(x,\lambda)}=-\bm\varSigma\varPsi\sb 3(x,\lambda),
\qquad
\overline{\varPsi\sb 4(x,\lambda)}=-\bm\varSigma\varPsi\sb 4(x,\lambda).
\end{eqnarray}

For
$\lambda\in(\lambda\sp\flat\sb{u},\lambda\sp\sharp\sb{u})$,
since
$\xi\sp\flat(\lambda)$ is real
and
$\xi\sp\sharp(\lambda)$ is imaginary,
and taking into account
\eqref{mpmm} and \eqref{pppm},
we see that
there are the relations
\begin{equation}\label{xixi}
\overline{\varXi\sp\flat\sb{+}(\lambda)e^{i\xi\sp\flat(\lambda) x}}
=\bm\varSigma\varXi\sp\flat\sb{-}(\lambda)e^{-i\xi\sp\flat(\lambda) x},
\qquad
\overline{\varXi\sp\sharp\sb{+}(\lambda)e^{i\xi\sp\sharp(\lambda) x}}
=\bm\varSigma\varXi\sp\sharp\sb{+}(\lambda)e^{i\xi\sp\sharp(\lambda) x}.
\end{equation}
Given the Jost solutions
$Y\sp\flat\sb\pm(x,\lambda)$
and
$Y\sp\sharp\sb\pm(x,\lambda)$
which satisfy
$(\eub{JL}-\lambda)\varPsi=0$,
with $\lambda\in i\R$,
we know that
$\bm\varSigma\overline{Y\sp\flat\sb\pm(x,\lambda)}$
and
$\bm\varSigma\overline{Y\sp\sharp\sb\pm(x,\lambda)}$
also
satisfy
$(\eub{JL}-\lambda)\varPsi=0$.
Matching the asymptotics
of the Jost solutions with \eqref{xixi}
(see Lemma~\ref{lemma-jost}),
we conclude that
\begin{equation}\label{yfs}
\overline{Y\sp\flat\sb{+}(x,\lambda)}
=\bm\varSigma Y\sp\flat\sb{-}(x,\lambda),
\qquad
\overline{Y\sp\sharp\sb{+}(x,\lambda)}
=\bm\varSigma Y\sp\sharp\sb{+}(x,\lambda).
\end{equation}
Taking into account
\eqref{vp13} and \eqref{yfs},
we have:
\[
\overline{E\sp{\blackdiamond}\sb{-+}}
=\det\left[
\overline{\varPsi\sb 1},\overline{\varPsi\sb 3},
\overline{Y\sp\flat\sb{-}},\overline{Y\sp\sharp\sb{+}}
\right]
=\det\left[
\bm\varSigma\varPsi\sb 1,-\bm\varSigma\varPsi\sb 3,\bm\varSigma Y\sp\flat\sb{+},\bm\varSigma Y\sp\sharp\sb{+}
\right]
=\det\left[
\varPsi\sb 1,-\varPsi\sb 3,Y\sp\flat\sb{+},Y\sp\sharp\sb{+}
\right]
=E\sp{\blackdiamond}\sb{++}.
\]
In the same manner one proves that
$\overline{
E\sp{\bullet}\sb{-+}(\lambda)}=E\sp{\bullet}\sb{++}(\lambda)$
for
$\lambda\in(\lambda\sp\flat\sb{u},\lambda\sp\sharp\sb{u})$.

The proof for
$\lambda\in(\lambda\sp\flat\sb{d},\lambda\sp\sharp\sb{d})$
is similar.
\end{proof}

\section{Conclusion}

We considered the spectrum of the nonlinear Dirac equation
in 1D,
linearized at a solitary wave solution.
The numeric simulations have been performed for
the nonlinearity $g(s)=1-s$
(the Soler model),
while some of our analytical conclusions
remain valid for any nonlinearity.

In particular,
we found that for any nonlinearity $g(s)$
there are the eigenvalues $\pm 2\omega i$
of the linearization $\eub{J}\eub{L}$.
For a certain range of $\omega$,
these eigenvalues
are embedded in the essential spectrum of $\eub{J}\eub{L}$.

For the
nonlinear Dirac equation with the nonlinearity $g(s)=1-s$
we have not found any other embedded eigenvalues
of $\eub{J}\eub{L}$.
We have not found any complex eigenvalues off the imaginary axis,
concluding that
the linearization
at all solitary waves
is spectrally stable.


\def\cprime{$'$} \def\cprime{$'$} \def\cprime{$'$} \def\cprime{$'$}
  \def\cprime{$'$} \def\cprime{$'$} \def\cprime{$'$} \def\cprime{$'$}
  \def\cprime{$'$} \def\cydot{\leavevmode\raise.4ex\hbox{.}} \def\cprime{$'$}
  \def\cydot{\leavevmode\raise.4ex\hbox{.}} \def\cprime{$'$} \def\cprime{$'$}
  \def\cprime{$'$} \def\cprime{$'$}

\end{document}